\documentclass{article}

\usepackage{arxiv}

\usepackage[utf8]{inputenc} 
\usepackage[T1]{fontenc}    
\usepackage{hyperref}       
\usepackage{url}            
\usepackage{booktabs}       
\usepackage{amsfonts}       
\usepackage{nicefrac}       
\usepackage{microtype}      
\usepackage{lipsum}
\usepackage{graphicx}
\graphicspath{ {./images/} }
\usepackage{amsmath}
\usepackage{lscape}
\usepackage{authblk}
\usepackage{caption}  

\title{TopoQA: a topological deep learning-based approach for protein complex structure interface quality assessment}

\author[1,2]{Bingqing Han}
\author[2]{Yipeng Zhang}
\author[2,3,4]{Longlong Li}
\author[1,*]{Xinqi Gong}
\author[2,*]{Kelin Xia}

\affil[1]{Institute for Mathematical Sciences, Renmin University of China, Beijing, China, 100872}
\affil[2]{Division of Mathematical Sciences, School of Physical and Mathematical Sciences, Nanyang Technological University, Singapore 637371, Singapore}
\affil[3]{School of Mathematics, Shandong University, Jinan 250100, China}
\affil[4]{Data Science Institute, Shandong University, Jinan 250100, China}

\affil[ ]{Emails: bingqinghan@ruc.edu.cn, yipeng001@e.ntu.edu.sg, longlee@mail.sdu.edu.cn, xinqigong@ruc.edu.cn, xiakelin@ntu.edu.sg} 

\begin{document}
\maketitle
\begin{abstract}
Even with the significant advances of AlphaFold-Multimer (AF-Multimer) and AlphaFold3 (AF3) in protein complex structure prediction, their accuracy is still not comparable with monomer structure prediction. Efficient and effective quality assessment (QA) or estimation of model accuracy (EMA) models that can evaluate the quality of the predicted protein-complexes without knowing their native structures, are of key importance for protein structure generation and model selection. In this paper, we leverage persistent homology (PH) to capture the atomic-level topological information around residues and design a topological deep learning-based QA method, TopoQA, to assess the accuracy of protein complex interfaces. We integrate PH from topological data analysis into graph neural networks (GNNs) to characterize complex higher-order structures that GNNs might overlook, enhancing the learning of the relationship between the topological structure of complex interfaces and quality scores. Our TopoQA model is extensively validated based on the two most-widely used benchmark datasets, DBM55-AF2 and HAF2, along with our newly constructed ABAG-AF3 dataset to facilitate comparisons with AF3. For all three datasets, TopoQA outperforms AF-Multimer-based AF2Rank and shows an advantage over AF3 in nearly half of the targets. In particular, in the DBM55-AF2 dataset, a ranking loss of 73.6\% lower than AF-Multimer-based AF2Rank is obtained. Further, other than AF-Multimer and AF3, we have also extensively compared with nearly-all the state-of-the-art models (as far as we know), it has been found that our TopoQA can achieve the highest Top 10 Hit-rate on the DBM55-AF2 dataset and the lowest ranking loss on the HAF2 dataset. Ablation experiments show that our topological features significantly improve the model's performance. At the same time, our method also provides a new paradigm for protein structure representation learning.
\end{abstract}


\section{Introduction}
The structures of protein complexes are of essential importance for the understanding of their molecular mechanisms, drug design and discovery, protein design, etc. Even though experimental methods can resolve protein 3D structures, they tend to be time-consuming and expensive \cite{jacobson2004comparative}, and not suitable for large-scale analysis. Data-driven models have been developed for protein 3D structure prediction \cite{alphafold2,alphafold-multimer,pro_pre1,pro_pre2,pro_pre3,pro_pre4,pro_pre5,baek2021accurate,all-atom}. Among them, AlphaFold2 \cite{alphafold2} has significantly advanced protein structure prediction, achieving performance in predicting monomer structures that rivals experimental methods. Recently, AlphaFold-Multimer (AF-Multimer) \cite{alphafold-multimer} and AlphaFold3 (AF3) \cite{alphafold3} have been developed for predicting protein complex structures. In particular, AF3 significantly improved the accuracy for antibody-antigen complex prediction \cite{alphafold3}. In AF3, a diffusion model-based framework is considered and various protein complex configurations are generated using different random seeds. When the native structures are absent, model quality assessment (QA) or estimation of model accuracy (EMA) is used in selection of the top-ranked configurations as the predicted structures. These QA and EMA models are critical for enhancing prediction reliability by estimating model quality in the absence of native structures \cite{2024recent}. In fact, EMA or QA methods are an important component of the Critical Assessment of Techniques for Protein Structure Prediction (CASP), which is a biennial experiment that advances and benchmarks protein structure prediction methods \cite{casp}, and were first introduced as a separate category since CASP7 \cite{assessment}. While EMA methods have evolved over the years, most of them are primarily designed for protein monomers \cite{QATEN,graphqa,deepqa,svmqa,voromqa}.

Mathematically, all EMA methods can be grouped into three categories\cite{2024recent}, including consensus models, pseudo-single models, and single models. Consensus methods assume near-native predicted structures are similar to each other, while poorly predicted structures differ greatly to each other \cite{MULTICOM_qa}. They assess model quality of a certain predicted structure through a pairwise comparison with all the other structures in the model pool, which is collection of all the generated structures from the same target sequence. The pairwise comparison is measured by scores like QS \cite{QS,openstructure}, lDDT \cite{lddt}, or DockQ \cite{dockq}, (with ModFOLDclust and MULTICOM\_qa as notable examples \cite{modfoldock,MULTICOM_qa}), the average of the scores is used to assess the quality of this structure. Consensus methods usually employ some well-established model pool. In contrast, pseudo-single model methods generate their own model pool for structure comparison. These two types of approaches are computationally expensive and their performance rely on the accuracy of the model pool \cite{2024recent}. For the single model methods, model pool is no longer required, thus they are do not have the above limitations. Single models can be divided into two categories: energy/statistical potential-based and deep learning-based. For the first type of method \cite{zrank2,goap,voromqa}, an energy function based on physico-chemical information is constructed or judgment is made through statistical results under a large amount of observational data. For the deep learning-based methods such as GNN-DOVE \cite{gnn-dove}, DProQA \cite{dproqa}, ComplexQA \cite{complexqa} and GraphGPSM \cite{liu2023estimating}, they usually represent protein structures as graphs, design amino-acid sequence, structural and physico-chemical features on nodes (and edges), and apply graph neural network models (GNNs). In QA approaches, proteins are often represented as graphs, with residues as nodes and contacts between residues as edges. Some models also consider atoms as nodes to provide a more detailed representation \cite{gnn-dove}. In general, GNN-based QA methods excel at propagating information across the entire graph, which helps capture global structural patterns and provides insights into the overall folding of protein molecules.
\begin{figure*} %
\centering
\includegraphics[width=\textwidth]{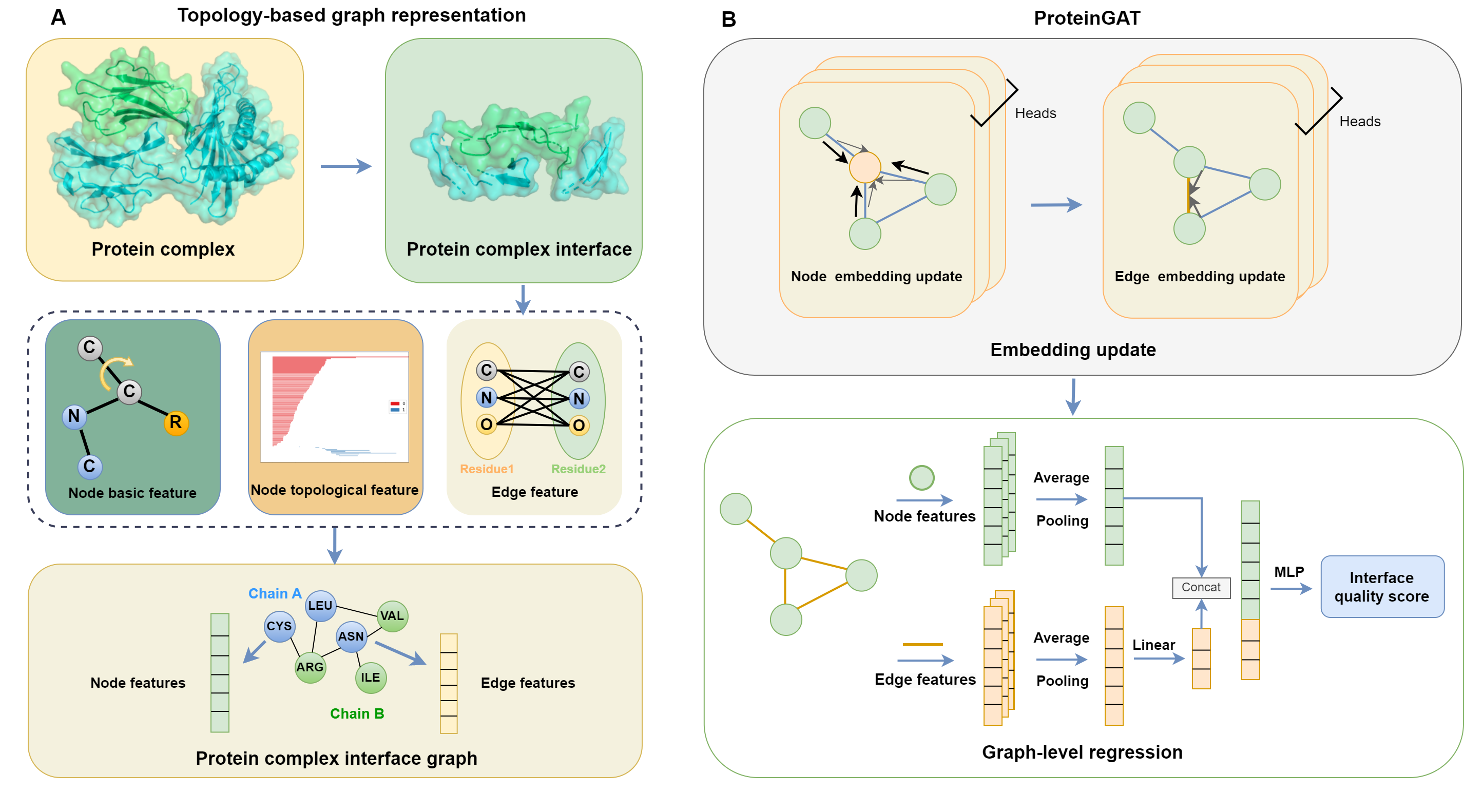}
\caption{TopoQA's architecture. (A) The process of generating the graph of the protein complex. (B) ProteinGAT module, divided into two parts: embedding update and graph-level regression prediction.}\label{topoqa}
\end{figure*}

Recently, topological data analysis and topological deep learning have been developed to explore high-order topological and geometric information within the data \cite{pun2022persistent,hofer2017deep,zia2024topological,papillon2023architectures,hajij2206topological,papamarkou2024position}. One effective approach is the integration of persistent homology (PH), which provides a robust mathematical framework for capturing and quantifying topological invariants across multiple scales. This enables the identification of complex, higher-order structures beyond traditional GNN models \cite{wangpersistent}. Such integration has demonstrated promising results across various domains, including biology \cite{xia2023persistent}, chemistry, physics, and image analysis \cite{TREPH}. PH can be incorporated into different GNN modules, such as feature representation \cite{horntopological, wong2021persistent, wangpersistent}, aggregation processes \cite{zhao2020persistence, TREPH}, pooling layers \cite{ying2024boosting}, and even loss functions \cite{ying2024boosting}. By combining PH with GNNs, models are better equipped to capture complex structures and show significant potential, particularly in predicting properties related to large biomolecules, such as proteins.

Here, we propose a topological deep learning-based model \textbf{TopoQA} for protein complex interface quality assessment for the first time. This model combined PH and GNN for protein structure representation learning. On the one hand, we simultaneously utilized the powerful learning ability of GNN and representation ability of PH to capture high-order local residue-level structural information; on the other hand, the local residue-level information is updated and aggregated to give the global representation through the message-passing module of graph neural networks. In local residue-level, we extract the atoms around each residue as a point cloud, generate a series of simplicial complexes according to the filtration process, calculate the barcodes, and vectorize them using their statistical properties as part of the initial node features. For edge features, in addition to $C_\alpha-C_\alpha$ distances, we also calculated the pairwise distances between all atoms of two residues, which facilitated a more detailed geometric representation of protein complex interface. In the global protein-complex level, we used a module called ProteinGAT to update node and edge embeddings, followed by pooling their information for interface quality score prediction. The results show that our method TopoQA is one of the state-of-the-art QA methods, showing outstanding performance across three benchmark datasets. Among these, DBM55-AF2 and HAF2 are the most-widely used benchmark datasets, while the newly generated ABAG-AF3 dataset by AF3 further validates the robustness and broad applicability of our approach. In three datasets, TopoQA showed advantages over AF-Multimer-based AF2Rank, especially in the DBM55-AF2 dataset, where it achieved a ranking loss 73.6\% lower compared to AF-Multimer-based AF2Rank. Compared with AF3's QA module, we have an advantage in nearly half of the targets. Compared with all the state-of-the art QA methods as far as we know, TopoQA achieves the highest Top 10 Hit-rate on the DBM55-AF2 dataset and the lowest ranking loss on the HAF2 dataset. Ablation experiments show that our topological features significantly improve the model's performance, with increases of 66.9\% and 3.4 times on the DBM55-AF2 and HAF2 datasets, respectively. Our approach provides a new paradigm for QA in terms of topology, facilitating better protein structure learning.
\section{Results}
\subsection{Graph representation for protein complex}
\begin{figure}[!ht]%
\centering
\includegraphics[width=0.7\textwidth]{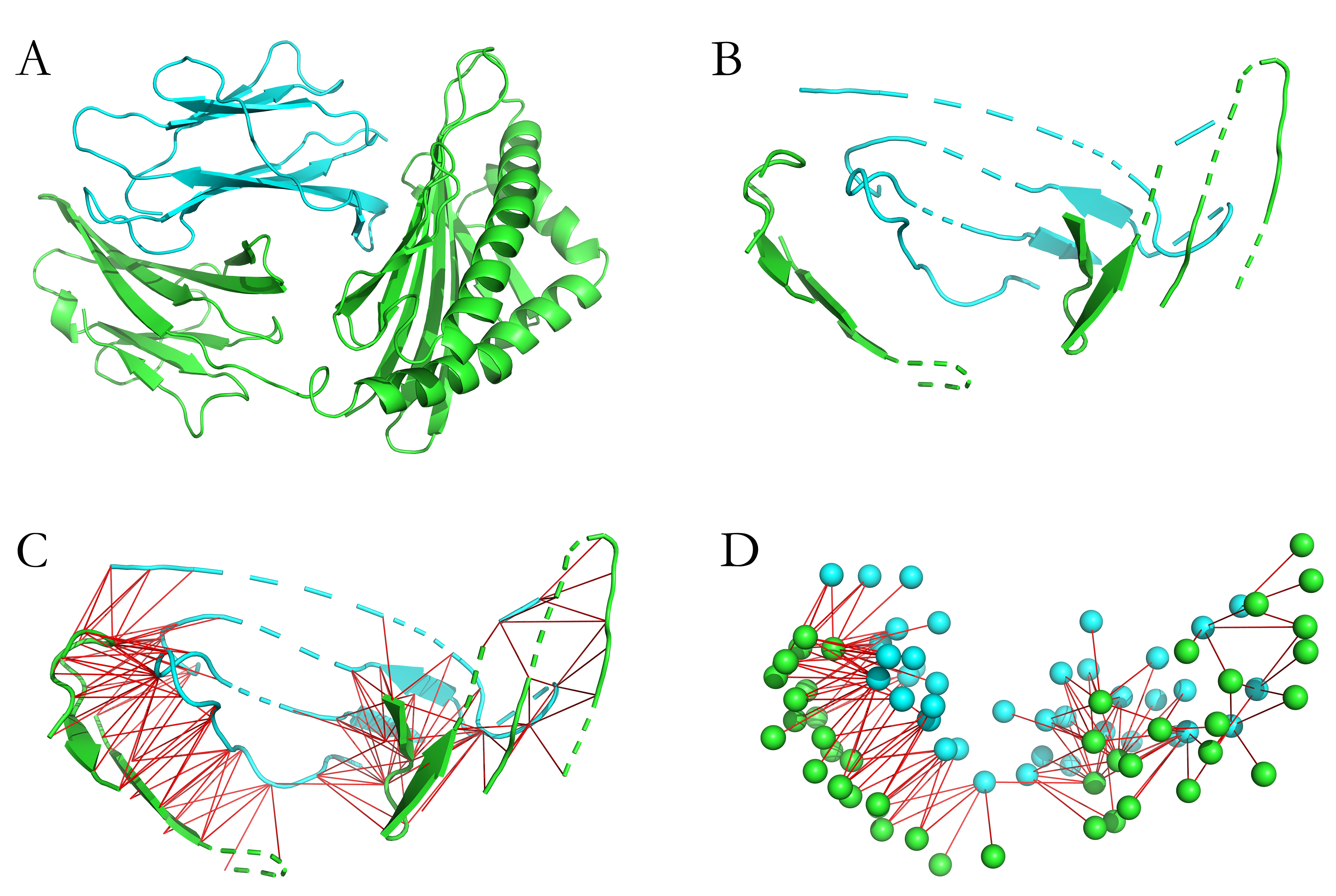}
\caption{(A) Complete protein complex with two chains. (B) Protein interface. (C) Inter-chain residue contacts on the original structure. (D) Inter-chain residue contacts on spheres drawn using $C_\alpha$ atom coordinates.}\label{fig1}
\end{figure}

\paragraph{Bipartite/Multipartite protein interface graph}
The topological representations of the protein complex can directly influence the performance of deep learning models. Here we propose bipartite/multipartite graph representation for the characterization of protein complex interface.  As shown in Figure~\ref{topoqa}A, since the interface is of key importance for protein complexes, we focus only on the interactions within interfaces and represent them as bipartite or multipartite graphs. Figure~\ref{fig1}A and B show the complete protein complex and the corresponding protein interface, respectively. Based on the extracted protein interface, we construct a bipartite interface graph $G=(V,E)$ to model the inter-chain interactions. As shown in Figure~\ref{fig1}C and D, in this bipartite graph, the residues are represented as vertices $V$ and the inter-chain contacts are represented as edges $E$. This bipartite structure allows us to effectively capture the complex interactions between different chains within the protein. Further, the detailed residue-level information is incorporated into our graph representation by considering special node features and edge features. 
\paragraph{Node featurization} 
Different from all previous models, we incorporate higher-order geometric and topological information of the local residual environments into our model by using the persistent homology analysis. More specifically, we compute PH on the local point cloud of each residue. We apply the element-specific skill \cite{wee2022persistent}: we extract the 3D coordinates of the $C_\alpha$ atom of target residue  and its surrounding atoms as point clouds, and divide the point cloud into different subsets according to the atom type carbon (C), nitrogen (N) and oxygen (O):\{\{C\},\{N\},\{O\},\{C,N\},\{C,O\},\{N,O\},\{C,N,O\}\}. We construct simplicial complexes using the above subsets of the point cloud. Using PH, the original point-cloud data is characterized by topological barcodes. We use five statistics for barcode vectorization: minimum, maximum, mean, sum and standard deviation. 
We compute the 0-dimensional and 1-dimensional barcodes and vectorized them into 140-dimensional topological features, which were added to the node features.

Computationally, each node has 172-dimensional features, including 32-dimensional basic features and 140-dimensional topological features.  Basic features include 21-dimensional one-hot encoding of residue types, 8-dimensional one-hot encoding of secondary structure types, 1-dimensional relative solvent accessible surface area, and 2-dimensional torsion angles. 

\paragraph{Edge featurization} 
The atomic interactions between adjacent residues within the protein interface are of key importance for the quality assessment of generated protein complexes.  To leverage detailed atomic information, a total of 10 edge features are based on atomic distances between the two residues. Specifically, for each edge connecting two residues, the atoms of each residue are divided into two point clouds. All pairwise distances between atoms in these point clouds are calculated and grouped into 10 bins. For each bin, the count of distances falling within that bin is used as the corresponding feature, capturing the distribution of atomic interactions between the residues. An extra edge feature for distance between $C_\alpha$ atoms is considered, resulting in 11-dimensional edge features.

\subsection{ProteinGAT}
We propose a special GNN architecture known as ProteinGAT. As shown in Figure~\ref{topoqa}B, our ProteinGAT model uses multi-head attention to update node and edge features, and leverage both to predict the interface quality score.

\subsubsection{Attention-based embedding update}
For the target node embedding $x_i^{(l)}$ at layer $l$, the attention coefficient $c_{ij}$ is computed using node embeddings $x_i^{(l)}$, $x_j^{(l)}$, and edge embedding $e_{ij}^{(l)}$, and then normalized into final weights $\alpha_{ij}$ via softmax function. The updated node embedding $x_i^{(l+1)}$ at layer $l+1$ is computed using attention-weighted information from neighboring nodes and itself. The updated edge embedding $e_{ij}^{(l+1)}$ at layer $l+1$ is obtained by concatenating the updated node embeddings $x_i^{(l+1)}$, $x_j^{(l+1)}$, and the original edge embedding $e_{ij}^{(l)}$, followed by a projection into a new vector space.
\subsubsection*{Graph-level regression}
After updating the embeddings, we apply average pooling to both node and edge embeddings. A linear layer reduces the pooled edge embeddings to half the dimension of the pooled node embeddings, highlighting the importance of node information. The concatenated embeddings are then passed through a multi-layer perceptron (MLP) for the final output.

We train the model using the Mean Squared Error (MSE) loss function, minimizing the difference between predicted values and DockQ scores.
\subsection{TopoQA model}
\subsubsection{Baselines}
We compared the performance of our model with two recently developed QA methods, ComplexQA \cite{complexqa} and DProQA \cite{dproqa}, both of which have demonstrated competitive performance in recent studies, using the same training, validation, and test sets for a fair comparison. In the blind CAPSP15 experiment, DProQA is one of the top performers among all single-model methods in terms of TM-score ranking loss \cite{chen2024survey}. Following previous work\cite{complexqa,dproqa}, we also include two deep learning methods, GNN-DOVE \cite{gnn-dove} and TRScore \cite{trscore}, as well as two classical energy-based methods, GOAP \cite{goap} and ZRANK2 \cite{zrank2}, all of which have shown high hit rates in their evaluations.

Specifically, referring to \cite{shuvo2023piqle}, we compared our method with the interface score (ipTM) predicted by AF-Multimer's self-assessment module. We utilized an extended version of the AF2Rank \cite{af2rank} method, which is based on the self-assessment module of AF-Multimer, repurposed for scoring protein complexes to generate the ipTM score. AF2Rank composite confidence score significantly outperformed all other EMA methods entered in casp14 \cite{af2rank}. We also use the ipTM of AF3 for comparison, which is one of the most advanced protein complex prediction models.
\subsubsection{Performance on three test datasets}
\begin{figure*}[!ht]%
\centering
\includegraphics[width=\textwidth]{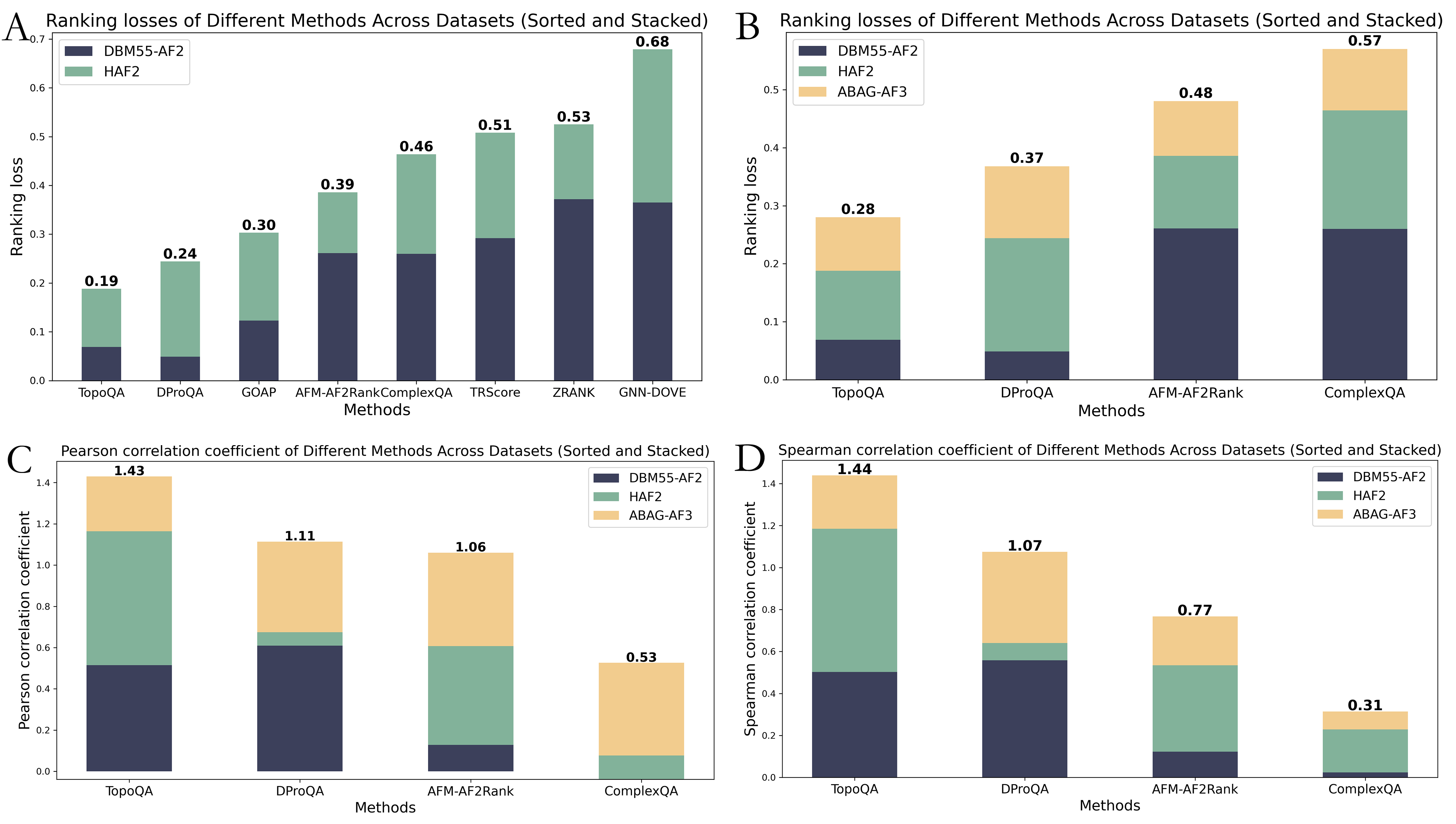}
\caption{Ranking losses and correlation coefficients of different methods across datasets (sorted and stacked). (A) Ranking losses of different methods across DBM55-AF2 and HAF2 datasets. (B) Ranking losses of different methods across DBM55-AF2, HAF2 and ABAG-AF3 datasets. (C) Pearson correlation coefficents of different methods aross DBM55-AF2, HAF2 and ABAG-AF3 dataset. (D) Spearman correlation coefficents of different methods aross DBM55-AF2, HAF2 and ABAG-AF3 dataset. AFM-AF2Rank refers to AF-Multimer-based AF2Rank.}\label{loss}
\end{figure*}

As shown in the Figure~\ref{loss}, we stacked the ranking losses from different datasets. In the DBM55-AF2 and HAF2 datasets, we followed previous work and used DockQ as the reference metric. For the ABAG-AF3 dataset, we used DockQ-wave as the reference metric, which can evaluate all interfaces of the complex. In Figure~\ref{loss}A, across the DBM55-AF2 and HAF2 datasets, TopoQA achieves the lowest stacked ranking loss among the eight methods. TopoQA's loss of 0.19 is 20.8\% lower than DProQA's second-best loss of 0.24, 36.7\% lower than GOAP's loss, and 51.3\% lower than AF-Multimer-based AF2Rank's loss. As shown in Figure~\ref{loss}B, TopoQA also achieves the lowest stacked ranking loss across the three datasets (DBM-55, HAF2 and ABAG-AF3). TopoQA's ranking loss is 0.28, which is 24.3\% lower than DProQA's loss of 0.37, and 41.7\% lower than AF-Multimer-based AF2Rank. Aggregating the results across different datasets, TopoQA demonstrates the best overall performance.

\paragraph{Performace on the DBM55-AF2 dataset}
Table S1 reports the ranking loss for all methods on the DBM55-AF2 dataset. Our method achieves the second lowest average ranking loss, is only 0.02 higher than the loss of DProQA with lowest loss 0.049. We achieved the ranking loss 0.069, is 43.9\% lower than GOAP with the third lowest ranking loss of 0.123, 73.5\% lower than ComplexQA with fourth lowest ranking loss of 0.26, and is 73.6\% lower than AF-Multimer-based AF2Rank with fifth lowest ranking loss of 0.261. For three targets 4ETQ, 5Y9J and 6AL0, TopoQA correctly selects the Top-1 model according to the DockQ, and achieves 0 ranking loss.

Table S2 shows the hit rate of different methods on the DBM55-AF2 dataset. TopoQA achieves the highest hit rate in all three level: selecting Acceptable or higher, Medium or higher and High-quality decoys. It is worth noting that TopoQA achieves the best possible Top-10 results at Medium and High quality levels.

As shown in Table S5 and S6, we also report the average results from multiple experiments with different random seeds. Compared to DProQA, TopoQA achieves a lower mean ranking loss and demonstrates more stable performance, with a standard deviation that is only 30.3\% of DProQA's. Additionally, according to Top-10 hit, TopoQA outperforms DProQA at the Acceptable level with 2.2 more targets.
\paragraph{Performace on the HAF2 dataset}
Table S3 reports the ranking loss for all methods on the HAF2 dataset. TopoQA achieves the lowest average ranking loss. TopoQA achieves the ranking loss 0.119, which is 4.8\% lower than AF-Multimer-based AF2Rank's second-lowest ranking loss of 0.125, 22.2\% lower than ZRANK2's third-lowest ranking loss of 0.153, 39.0\% lower than DProQA's ranking loss of 0.195. Moreover, TopoQA achieves the lowest loss on two targets 7AMV and 7D3Y. 

Table S4 shows the hit rate of different methods on the HAF2 dataset. TopoQA achieves the highest hit rate at two level: selecting Medium or higher-quality and High-quality decoys. At the Acceptable level, TopoQA's hit rate is 10, which is slightly lower than the best result of 11 achieved by other methods. TopoQA achieves the best possible Top-10 results at High quality level.

As shown in Table S7 and S8, on the HAF2 dataset, TopoQA outperforms DProQA with a 48.5\% lower mean and 70\% lower standard deviation in ranking losses. Additionally, TopoQA has a better average hit rate: at both Acceptable and Medium levels, it has 1.4 more targets on average than DProQA.
\paragraph{Performance on the ABAG-AF3 dataset}
Table~\ref{tab:AF3} reports the ranking loss and top 10 mean DockQ-wave for 5 methods on the ABAG-AF3 dataset. Except for AF3, TopoQA achieved the lowest ranking loss and the highest Top10 average DockQ-wave value, demonstrating the advantage of our method: it can select the better conformation from different conformations of the same target. We achieved the ranking loss 0.092, is 13.2\% lower than ComplexQA with the ranking loss of 0.106, and 25.8\% lower than DProQA with the ranking loss of 0.124. It is worth noting that, on this dataset, TopoQA also demonstrates better performance than AF-Multimer-based AF2Rank.

Although TopoQA does not outperform AF3 overall, as shown in Tables S9 and S10, it achieves lower ranking loss on 17 targets and better top-10 mean DockQ-wave scores on 16, out of a total of 35 targets. Although our training data is significantly smaller than that of AF3, TopoQA still shows advantages on nearly half of the targets, demonstrating the potential of the TopoQA model.
\begin{table}[!h]
\centering
\caption{Performance on ABAG-AF3 dataset using DockQ-wave as reference metric. AFM-AF2Rank refers to AF-Multimer-based AF2Rank. For the same target, there may be multiple top 1 models, and we select the average of the losses of these models as the ranking loss of the target.}
\label{tab:AF3}
\begin{tabular*}{0.8\textwidth}{@{\extracolsep{\fill}}lll}
\toprule
Method    & Ranking loss$\downarrow$ & Top10 mean$\uparrow$ \\
\midrule
TopoQA    & 0.092        & 0.592                 \\
AF3       & 0.054        & 0.614                 \\
DproQA    & 0.124        & 0.585                 \\
ComplexQA & 0.106        & 0.590                  \\
AFM-AF2Rank       & 0.094        & 0.589               \\ 
\bottomrule
\end{tabular*}
\end{table}
\subsubsection{Correlation coefficient analysis}
As shown in the Figure~\ref{loss}, we stacked the correlation coefficients from three datasets. In Figure~\ref{loss}C, TopoQA achieves the highest stacked pearson correlation coefficient, with a value of 1.43, which is 28.83\% higher than the second-best correlation coefficient of 1.11 achieved by DProQA. In Figure~\ref{loss}D, TopoQA also achieves the highest stacked spearman correlation coefficient, measured at 1.44, representing a 34.6\% increase compared to the second-best correlation coefficient of 1.07. Overall, TopoQA also shows good performance under the correlation coefficient evaluation metrics.
\subsubsection{Evaluation using different reference metrics}
In the DBM55-AF2 and HAF2 dataset, we primarily used DockQ and CAPRI criteria to evaluate models, following previous work. These metrics assess a single protein interface, while more recent metrics like QS-score \cite{QS}, DockQ-wave \cite{assessment} and BDM \cite{bdm} evaluate all interfaces. As shown in Table~\ref{tab:metric}, TopoQA performed consistently well across different metrics, with ranking losses of 0.069 (DockQ), 0.063 (DockQ-wave), and 0.044 (QS-score). Additionally, correlation coefficients were all above 0.5, demonstrating the robustness and stability of TopoQA in evaluating interface accuracy.
\begin{table}[h]
\centering
\caption{Performance of TopoQA under different reference metrics on the DBM55-AF2 dataset.DockQ-Wave and QS-score values were calculated by OpenStructure\cite{openstructure}.}
\label{tab:metric}  
\begin{tabular*}{0.8\textwidth}{@{\extracolsep{\fill}}llll}
\toprule
Reference metrics & Ranking loss & PearsonCor & SpearCor \\
\midrule
DockQ             & 0.069        & 0.515      & 0.502    \\
DockQ-wave        & 0.063        & 0.506      & 0.541    \\
QS-score          & 0.044        & 0.546      & 0.512    \\
\bottomrule
\end{tabular*}
\end{table}

\subsubsection{Ablation study}
To evaluate the impact of node topological and atomic distance-related edge features, we performed ablation studies by removing specific components of the TopoQA model, with the results shown in Table S11 and Figure S1.
\paragraph{The impact of the node topological features}
We removed the topological features of the nodes and only kept the basic features. The results showed that the performance of the model was greatly affected. Specifically, on the DBM55-AF2 dataset, the ranking loss worsened from 0.069 to 0.129, representing an 87.0\% increase; the Pearson correlation coefficient and Spearman correlation coefficient decreased by 0.198 (a 38.4\% reduction) and 0.122 (a 24.3\% reduction), respectively. On the HAF2 dataset, the ranking loss worsened from 0.119 to 0.157, indicating a 31.9\% increase, while the Pearson and Spearman correlation coefficients decreased by 0.482 (a 74.3\% reduction) and 0.417 (a 61.1\% reduction), respectively.

\paragraph{The impact of the edge features related to all atomic distances}
We removed the atomic distance-related edge features, keeping only the $C_\alpha-C_\alpha$ distance. This led to a decline in model performance. On the DBM55-AF2 dataset, the ranking loss worsened from 0.069 to 0.103 (a 49.3\% increase), with the Pearson correlation coefficient increased by 0.01, but the Spearman coefficient dropping by 0.011 (2.2\%). On the HAF2 dataset, the ranking loss worsened from 0.119 to 0.165 (38.7\%), and both Pearson and Spearman coefficients decreased by 0.02 (3.1\%) and 0.049 (7.2\%), respectively.

\setlength{\parskip}{1em} 
The results show that node topological features and edge features related to all atomic distances both have an effect on improving model performance. Among them, as shown in the Figure S1, the node topological features have the significant impact, after removing topological features, the sum of the model's metrics (with ranking loss negated) on the DBM55-AF2 and HAF2 datasets is 59.9\% and 22.8\% of TopoQA, respectively. It shows the powerful ability of combining PH and GNN and its great potential in protein structure learning.
\section{Discussion}
AF-Multimer and AF3 have been developed for protein complex structure prediction, but compared to monomer structure prediction, there is still room for improvement in accuracy. We propose a topology-based quality assessment method to enhance model selection and improve the accuracy of protein complex structure predictions.

In previous studies, residues were commonly used as nodes to construct graphs for GNN-based prediction, which helps capture global structural patterns and provides insights into the overall folding of protein molecules. We integrate persistent homology with GNNs for quality assessment, using PH to capture atomic-level topological information around the target residue, thus enhancing the model's representation of protein structures. Ablation studies demonstrate that incorporating PH significantly improves model performance. Our method shows great potential and can be extended to other protein structure representation tasks.

EMA or QA methods are an important part of CASP experiments. In the recent CASP15 EMA, the task involved 40 targets and 10329 models \cite{assessment}, including complexes ranging from dimers to 27 chains. Some teams in CASP EMA, such as GuijunLab-RocketX \cite{liu2023estimating} and Chaepred, trained on large datasets with up to 2 million and 400000 decoys, respectively. In contrast, our model was trained on a smaller dataset of 8,733 decoys, primarily oligomeric proteins. We believe that with a larger dataset, TopoQA could demonstrate greater potential.

We compared our method, TopoQA, with the AF-Multimer-based AF2Rank and AF3 across three datasets. TopoQA outperformed AF-Multimer-based AF2Rank on all datasets, particularly on the DBM55-AF2 dataset, where our ranking loss was 73.5\% lower. Notably, we trained TopoQA using only a small dataset generated by AF-Multimer, which is significantly less than the training data used for AF-Multimer. Although the overall performance of TopoQA is not as high as that of AF3, it demonstrates advantages on nearly half of the targets, indicating its potential. Furthermore, AF3 currently provides quality scores only for its predicted structures, while our method offers greater applicability and versatility for model quality assessment.

Currently, our model, TopoQA, is designed to assess the global interface accuracy of protein complexes. EMA methods, however, also encompass evalutions of global fold and local accuracy. Global fold accuracy focuses on the overall correctness of the complex structure, utilizing metrics such as TM-score and GDT-score \cite{gdt} to evaluate global topology. In contrast, local accuracy assesses the residue-level precision, employing metrics like lDDT \cite{lddt} and CAD-score \cite{cad} as reference values. In the future,we plan to incorporate multi-task learning to broaden our model’s capabilities from interface accuracy evaluation to a more comprehensive assessment of accuracy.
\section{Conclusion}
In this work, we present a topological deep learning-based method, TopoQA, for protein complex structure interface quality assessment. Constructing graphs with residues as nodes combined with GNNs is a common approach, and can allow for modeling complex interactions between residues, thereby capturing important global structural patterns. We enhance protein structural representation by incorporating topological information using PH. For each residue, we extract the topological features of its neighboring atoms based on specific atomic combinations. Ablation experiments reveal that the introduced topological features significantly enhance the model. When these features are removed, the model's performance drops to 59.9\% and 22.8\% of its original level on two datasets, respectively. 

Compared to other models, our method achieves highest hit rate on the DBM55-AF2 dataset and lowest rank loss on the HAF2 dataset. Additionally, multiple experiments with different random seeds and evaluation reference metrics show that TopoQA produces stable results, demonstrating its robustness. On three datasets, our model all demonstrates advantages over AF-Multimer-based AF2Rank, particularly on the DBM-55 dataset, where TopoQA’s loss is 73.6\% lower than that of AF-Multimer. Additionally, compared to AF3's QA module, TopoQA shows an advantage on nearly half of the targets.

\section{Materials and methods}
\subsection{Datasets}
\subsubsection{Training and validation datasets}
We used the same training and validation sets as DProQA and ComplexQA. They combined the two datasets and divided them into training and validation sets. \textbf{Multimer-AF2 Dataset}: The MAF2 dataset comprises complex structures predicted by AlphaFold2 and AF-Multimer, with protein complex targets sourced from the EVCoupling \cite{hopf2019evcouplings} and DeepHomo \cite{deephomo} datasets. The MAF2 dataset contains 9251 decoys. \textbf{Dockground Dataset}: the Dockground Dataset\cite{dockground} contains 58 protein complex targets, each with averages of 9.83 correct and 98.5 incorrect decoys.
\subsubsection{Test datasets}
We used the following three test datasets to test our model. \textbf{Docking Benchmark5.5 AF2} (DBM55-AF2) Dataset, comprises 15 antibody-antigen complex targets and 449 decoy models. \textbf{Heterodimer-AF2} (HAF2) Dataset is also generated by AF-Multimer, and contains 13 heterodimer targets with 1370 decoy models. \textbf{ABAG-Docking Benchmark AF3} (ABAG-AF3) Dataset: In our previous work \cite{zhao2024abag}, we compiled a non-redundant antibody-antigen dataset. We selected proteins released after 2022 as targets and used AF3 to generate 25 conformations per target, running it five times with different seeds. The ABAG-AF3 dataset consists of 35 targets and 875 conformations.
\subsection{TopoQA}
\subsubsection{Persistent Homology}
\textbf{Simplicial Complex:} A \textbf{simplicial complex} is a collection of simplices (geometric objects such as points, edges, triangles, and their higher-dimensional counterparts) that are combined in a way that preserves their geometric structure. A \textbf{$k$-simplex} $\sigma_k$, the fundamental building block of a simplicial complex, is defined as the convex hull of $k+1$ affinely independent points $v_0, v_1, \dots, v_k$ in $\mathbb{R}^N$:
\[
\sigma_k = \left\{ \sum_{i=0}^{k} \lambda_i v_i \mid \lambda_i \geq 0, \sum_{i=0}^{k} \lambda_i = 1 \right\}.
\]
A 0-simplex is a vertex, a 1-simplex is an edge, a 2-simplex is a triangle, and a 3-simplex is a tetrahedron. The dimension of a simplex is the number of vertices minus one.

A simplicial complex encodes richer, higher-dimensional information than a graph, making it an ideal framework for describing the shape and structure of complex objects. Its ability to capture relationships beyond pairwise connections allows for a deeper analysis of the object's topological and geometric properties.

\textbf{Homology Group:} Homology groups are algebraic structures that capture topological invariants of a simplicial complex, providing information about its shape and structure. Specifically, they describe features such as connected components, loops, and voids in different dimensions. These groups are crucial in distinguishing spaces with similar local structures but different global topological properties, making them a powerful tool for analyzing the intrinsic characteristics of the simplicial complex.

For a given simplicial complex $K$, a $k$-chain is a formal sum of $k$-simplices with coefficients from a field, typically $\mathbb{Z}_2$. The set of all $k$-chains forms an Abelian group, denoted as $C_k(K; \mathbb{Z}_2)$.

The boundary operator $\partial_k : C_k \to C_{k-1}$ maps each $k$-simplex $\sigma_k = [v_0, v_1, \dots, v_k]$ to its $(k-1)$-dimensional faces:
\[
\partial_k \sigma_k = \sum_{i=0}^{k} (-1)^i [v_0, v_1, \dots, \hat{v_i}, \dots, v_k],
\]
$[v_0, \dots, \hat{v_i}, \dots, v_k]$ represents the $(k-1)$-simplex obtained by omitting the vertex $v_i$ from the simplex $\sigma_k$. A key property is that applying the boundary operator twice results in zero: $\partial_{k-1} \circ \partial_k = 0$.

This allows us to define the cycle group $Z_k = \ker(\partial_k)$ (elements with no boundary) and the boundary group $B_k = \mathrm{im}(\partial_{k+1})$ (boundaries of higher-dimensional simplices). The $k$-th homology group is then defined as the quotient:
\[
H_k(K; \mathbb{Z}_2) = Z_k / B_k.
\]
The rank of $H_k$, known as the betti number $\beta_k$, represents the number of $k$-dimensional holes in the simplicial complex.

\textbf{Persistent Homology:} While classical homology captures topological features of a space, it does not contain geometry information like the scale of the object. Persistent homology addresses this by tracking the appearance and disappearance of homology classes over a filtration, providing additional geometric and scale-related information. This makes it a powerful tool for analyzing shapes, capturing features that persist across multiple scales.

For a simplicial complex $K$, a filtration is a sequence of nested subcomplexes:
\[
\emptyset = K_0 \subseteq K_1 \subseteq \dots \subseteq K_m = K.
\]
As the filtration progresses, topological features (like connected components, loops, and voids) are created and eventually disappear. Persistent homology tracks these features over the filtration.

The \textbf{$p$-persistent $k$-th homology group} is defined as:
\[
H_k^p(K_i) = Z_k(K_i) / (B_k(K_{i+p}) \cap Z_k(K_i)).
\]
This measures how long a homology class persists across different filtration levels.

To use persistent homology as a feature, we track the birth time and death time of each generator in the persistent homology groups. The birth time marks the filtration level at which a generator first appears, while the death time indicates when it either merges with another generator or vanishes. These times provide valuable insights into the persistence of topological features across different scales. 
\subsubsection{Graph representation for protein complex interface}
To assess the interface quality, we retained interface residues within 10Å of the $C_\alpha$ atom of at any residue in the other chain. To focus on inter-chain interactions, we only consider inter-chain edges, defined as connections between residues from different chains with $C_\alpha-C_\alpha$ distances less than 10Å. We then construct the interface graph $G(V,E)$, where residues (represented by their $C_\alpha$ atoms) as the vertices $V$ and the inter-chain contacts are the edges $E$. 
\subsubsection{Node topological features}
For each residue, we use PH to extract its topological information. Specifically, for residue $i$, we consider the 3D coordinates set $A_{i}$, which includes the coordinates of neighboring atoms within a cut-off distance $r$ (we choose 8Å here) from the $C_\alpha$ atom of the target residue. We apply the element-specific skill: we divide the point cloud $A_{i}$ into different subsets $A_{i\_specific}$ according to the atom type:\{\{C\},\{N\},\{O\},\{C,N\},\{C,O\},\{N,O\},\{C,N,O\}\}. For each point cloud $A_{i\_specific}$ with each atom type, we apply the \textbf{Vietoris-Rips complex} for calculating 0-dimensional PH, which is a simplicial complex generated from a set of points $X$ in a metric space by connecting points with edges if their pairwise distances are below a threshold $\epsilon$. Specifically, a $k$-simplex is formed if every pair of its $k+1$ vertices is at most $\epsilon$ apart. $\epsilon$ is the filtration value.

And we apply the \textbf{Alpha complex} for calculating 1-dimensional PH. Given a set of points $X$ in a metric space and a radius parameter $\alpha$, the Alpha complex is a simplicial complex constructed as follows: a $k$-simplex is included if and only if its $k+1$ vertices can be enclosed by a ball of radius $\alpha$ that contains no other points from $X$ inside or on the boundary of the ball, apart from the $k+1$ vertices themselves. 

Using the given simplicial complexes, we construct a filtration of simplicial complexes and compute the associated PH and barcodes. A \textbf{barcode} is a visual representation in which each bar corresponds to a specific generator of PH groups, where generators in different dimensions represent distinct topological features, such as connected components in dimension 0, loops in dimension 1, and voids in higher dimensions. The left endpoint of a bar marks the \textit{birth} of a generator, while the right endpoint marks its \textit{death}. The length of the bar, representing the difference between the birth and death values, quantifies the \textit{persistence} of the generator, providing insight into its significance within the underlying topological space.

To get a fixed-sized feature vector from the persistent homology barcodes, we consider five statistics: average (avg), standard deviation (std), maximum (max), minimum (min), sum (sum). For 0-dimensional barcode, since \textit{birth} is always 0, we only use death for calculation. For 1-dimensional barcode, we take \textit{birth}, \textit{death} and $death-birth$ (persistence of bars) for statistic calculation. We also filtered the barcodes: bars with late death times in 0-dimensional barcode; bars with very short lifetimes: these bars are typically considered noise in the data. We use the following criteria to filter:
\begin{equation}
death\leq8
\end{equation}
\begin{equation}
death-birth\geq0.01
\end{equation}

In summary, for each residue, our methodology yields a feature vector with a dimensionality of  $140 = 7 \times 5 \, (\text{statistics}) (\text{0-dimensional barcode}) + 7 \times 5 \, (\text{statistics}) \times 3 \, (\text{birth, death, and persistence of bars}) \, (\text{1-dimensional}$ $\text{barcode})$, providing a robust foundation for subsequent quality prediction tasks.

\subsubsection{Edge features}
We added an 11-dimensional edge feature to the edge formed between two residues, where the first dimension represents the distance between their $C_\alpha$ atoms, and the remaining 10 dimensions capture atomic distances between the two residues. Specifically, these 10 features are derived by constructing a bipartite graph between two point clouds, each point cloud representing the atoms of the corresponding residues. We computed all pairwise distances between atoms in the two point clouds and divided the interval $[1, +\infty]$ into 10 bins: $[1,2)$, $[2,3)$, $[3,4)$, $[4,5)$, $[5,6)$, $[6,7)$, $[7,8)$, $[8,9)$, $[9,10)$, and $[10, +\infty)$. For each bin, we constructed a bipartite graph using only the edges corresponding to distances within that bin, and the count of edges in each bipartite graph was used as the feature for the corresponding dimension of the edge connecting the two residues.

\subsubsection{ProteinGAT module}\label{subsec4}
ProteinGAT module is designed to update node and edge embeddings based on multi-head attention, and perform graph-level regression prediction.

\textbf{Embedding update}. We use a multi-head attention mechanism. Below, we take the calculation of one head as an example to illustrate the calculation process. Assume that the node embedding and edge embedding at layer $l$ are $\mathbf{x}_i^{(l)}$ and $\mathbf{e}_{ij}^{(j)}$ respectively, the node and edge embedding at layer $l+1$ are $\mathbf{x}_i^{(l+1)}$, $\mathbf{e}_{ij}^{(l+1)}$, respectively. The update formulas are:
\begin{equation}
c_{ij}=\sigma(\mathbf{W}_s\mathbf{x}_i+\mathbf{W}_t\mathbf{x}_j+\mathbf{W}_ee_{ij})
\label{eq:att}
\end{equation}
\begin{equation}
\alpha_{ij}=\frac{exp(c_{ij})}{\sum_{k\in\mathcal{N}(i)\cup\{i\}}exp(c_{ik})}
\label{eq:att_normal}
\end{equation}
\begin{equation}
\mathbf{x}_i^{(l+1)}=\sum_{j\in\mathcal{N}(i)\cup\{i\}}\alpha_{ij}\mathbf{\Theta}_t\mathbf{x}_j^{(l)}
\label{eq:node_update}
\end{equation}
\begin{equation}
\mathbf{e}_{ij}^{(l+1)}=\mathbf{\Theta}_e(\mathbf{x}_i^{(l+1)}||\mathbf{x}_j^{(l+1)}||\mathbf{e}_{ij}^{(l)})
\label{eq:edge_update}
\end{equation}
Equation~\eqref{eq:att} computes the attention coefficient between node $i$ and node $j$. The trainable weight matrices $\mathbf{W}_s$, $\mathbf{W}_t$ and $\mathbf{W}_e$ map the source node, end node and edge features to the new feature space respectively. $\sigma$ is the activation function. Equation~\eqref{eq:att_normal} uses the softmax function to normalize the attention coefficients $c_{ij}$ to obtain the final weights $\alpha_{ij}$. $\mathcal{N}(i)$ is the set of the neighbor nodes of node $i$. Equation~\eqref{eq:node_update} updates node embedding, where $\mathbf{\Theta}_t$ is the trainable weight. Equation~\eqref{eq:edge_update} uses the node embeddings from layer $(l+1)$ and the edge embeddings from layer $l$ to update the edge embeddings for layer $(l+1)$, where $\mathbf{\Theta}_e$ represents the trainable weight matrix and '$||$' denotes the vector concatenation operation.

\textbf{Graph-level regression}. In graph-level prediction, we use both node and edge information. Assume that final node embedding and edge embedding are $\mathbf{x}_i^{(L)}$ and $\mathbf{e}_{ij}^{(L)}$, respectively, the node features and edge features after pooling are $\mathbf{x}_{pool}$ and $\mathbf{e}_{pool}$:
\begin{equation}
\mathbf{x}_{pool}=\frac{1}{|V|}\sum_{i\in V}\mathbf{x}_i^{(L)}
\end{equation}
\begin{equation}
\mathbf{e}_{pool}=\mathbf{W}_{pool}\frac{1}{|E|}\sum_{ij\in E}\mathbf{e}_{ij}^{(L)}
\end{equation}
After $L$ updates, we applied average pooling to the embeddings of both nodes and edges. For the edge embeddings, we added a linear layer to reduce their dimension to half of that of the node embeddings, emphasizing the greater importance of node information compared to edge features. We integrated both node and edge information, and fed the concatenated vector into a MLP with three stacked linear layers for the final out put $o$:
\begin{equation}
    o=Sigmoid(MLP(\mathbf{x}_{pool}||\mathbf{e}_{pool}))
\end{equation}
During training, the model is optimized to minimize the MSE loss.

\section*{Acknowledgement}{This work was supported in part by Singapore Ministry of Education Academic Research fund Tier 1 grant RG16/23, Tier 2 grants MOE-T2EP20120-0010 and MOE-T2EP20221-0003; Program of China Scholarship Council (Grant No.202306360241), Interdisciplinary Innovative Research Program of School of Interdisciplinary Studies, Renmin University of China. It is also supported by Public Computing Cloud, Renmin University of China.}

\bibliographystyle{unsrt}  





\bibliography{references} 

\appendix  
\section{Supporting text}
\subsubsection*{Visualization analysis}
To intuitively illustrate the performance of TopoQA in classifying decoys, we use Principal Component Analysis (PCA) \cite{pca} and t-distributed Stochastic Neighbor Embedding (t-SNE) \cite{t-sne} to reduce the dimensions of the encodings generated by TopoQA for visualization. PCA is a linear dimension-reduction method that projects data into a new coordinate system through orthogonal transformation. t-SNE is a nonlinear dimension-reduction method that preserves the local structure and similarity of data by minimizing the Kullback-Leibler divergence. We extracted the feature embedding vectors from the penultimate layer of the model as input data for dimension-reduction. By mapping high-dimensional features into a two-dimensional space, we can observe the distribution of Acceptable or higher quality and Incorrect-quality samples in the lower-dimensional space. As shown in the Figure~\ref{keshihua}, we applied the above two dimension-reduction techniques to the two test sets DBM55-AF2 and HAF2. Acceptable decoys and Incorrect decoys are relatively separated in low-dimensional space, and TopoQA can distinguish between these two types of decoys in most cases. The visualization results show that TopoQA has strong classification and generalization capabilities on unseen data, and the learned feature representation can effectively capture the relationship between structure and quality.

\subsection*{Methods}
\subsubsection*{Normalization for features}\label{subsubsec3}
The feature scales we selected are different. In order to improve the stability of the model and avoid the impact of scale differences on the model, we normalized the features of nodes and edges. We use the Min-Max Normaliztion, assume that the $i-th$ (node/edge) feature in graph $G$ is $fea_i$, and the normalized feature is $normalized\_fea_i$:
\begin{equation}
normalized\_fea_i=\frac{fea_i-min(fea_i)}{max(fea_i)-min(fea_i)}
\end{equation}

\subsubsection*{Evaluation metrics}\label{subsec5}
Evaluation metrics can be divided into reference metrics and statistical metrics. Reference metrics assess the accuracy of structural models, while statistical metrics, such as ranking loss, evaluate the ability of QA methods to predict these reference metrics. We utilized the following \textbf{reference metrics}:

\textbf{DockQ} \cite{dockq} combines three interface similarity metrics: L-RMSD, the $C_\alpha-RMSD$ of the ligand in the model relative to the reference structure; I-RMSD, the $C_\alpha-RMSD$ of the interface region between the decoy model and the reference structure; and $F_{nat}$, the fraction of atom pairs correctly predicted in the decoy models. DockQ is a continuous value of $[0,1]$, and the larger the value, the higher the interface quality.

\textbf{CAPRI criteria} \cite{capri} combines L-RMSD, I-RMSD and $F_{nat}$ to classify predicted structures into four levels: High-quality, Medium-quality, Acceptable-quality, and Incorrect.

\textbf{DockQ-wave} \cite{assessment} is a variation of DockQ. It is obtained by weighting the DockQ score of each interface.

\textbf{QS-score}\cite{QS} represents the fraction of shared interface contacts (residues on different chains with $C_\beta-C_\beta \text{ distance} < 12\text{\AA}$) between two structures. The range of QS-score is $[0,1]$, and when QS-score is close to 1, it means that the interfaces are very similar.

For decoy $i$, let the predicted quality score be $y_i$, and the reference value be $x_i$. We used the following \textbf{statistical metrics}.

\textbf{Pearson correlation coefficient} $PearsonCor$ is used to measure the linear relationship between the predicted quality value and the reference value:
\begin{equation}
    PearsonCor=\frac{\sum_{i=1}^n(x_i-\overline{x})(y_i-\overline{y})}{\sqrt{\sum_{i=1}^n(x_i-\overline{x})^2\sum_{i=1}^n(y_i-\overline{y})^2}}
\end{equation}

\textbf{Spearman correlation coefficient} $SpearCor$ is used to measure the monotonic relationship between two variables: 
\begin{equation}
SpearCor=1-\frac{6\sum_{i=1}^n d_i^2}{n(n^2-1)}
\end{equation}
\begin{equation}
d_i=R(x_i)-R(y_i)
\end{equation}
$R(x_i)$ and $R(y_i)$ are the ranks of $x_i$ and $y_i$.

\textbf{Ranking loss} is used to measure the ability of the QA models to correctly select the Top 1 model. Ranking loss is the difference between the highest reference value and the reference value corresponding to the Top 1 decoy selected by the QA methods.

\textbf{Top-10 hits rate} is represented by three numbers separated by the character/. These three numbers, in order, represent how many decoys with Acceptable or higher-quality, Medium or higher-quality, and High-quality are among the Top-10 ranked decoys.

\begin{landscape}
\section{Tables and Figures}

\renewcommand{\thetable}{S\arabic{table}}
\setcounter{table}{0}

\begin{table*}[!htbp]
\centering
\small
\caption{The DockQ score ranking loss results on DBM55-AF2 dataset.The final row reports the mean and standard deviation (Std) of the ranking loss of different methods. Results for the best-performing methods are shown in bold, and results for the second-best methods are underlined. AFM-AF2Rank refers to AF-Multimer-based AF2Rank.}
\label{tab:bm55_loss}  
\begin{tabular*}{1.2\textwidth}{@{\extracolsep{\fill}}lllllllll@{\extracolsep{\fill}}}

\toprule
Target  & TopoQA & ComplexQA\cite{complexqa} & DProQA\cite{dproqa} & AFM-AF2Rank & GNN-DOVE\cite{complexqa} & GOAP\cite{complexqa}  & ZRANK2\cite{complexqa} & TRScore\cite{complexqa} \\
\midrule
3SE8 & 0.102 & 0.663 & 0.079 & 0.058 & 0.663 & 0     & 0.735 & 0.408 \\
3U7Y & 0.021 & 0.021 & 0     & 0.758 & 0.756 & 0     & 0.772 & 0.745 \\
3WD5 & 0.011 & 0.138 & 0.011 & 0.683 & 0.672 & 0.011 & 0.704 & 0.105 \\
4ETQ & 0     & 0.748 & 0     & 0.755 & 0.759 & 0     & 0.759 & 0.759 \\
4M5Z & 0.269 & 0.015 & 0.015 & 0.015 & 0.211 & 0.133 & 0.221 & 0.133 \\
5CBA & 0.008 & 0.038 & 0.052 & 0.065 & 0.019 & 0.007 & 0.047 & 0.038 \\
5GRJ & 0.024 & 0.595 & 0.024 & 0.226 & 0.742 & 0.23  & 0.774 & 0.355 \\
5HGG & 0.079 & 0.051 & 0.051 & 0.04  & 0.033 & 0.051 & 0.051 & 0.072 \\
5KOV & 0.06  & 0     & 0.065 & 0.06  & 0     & 0.078 & 0.08  & 0.07  \\
5WK3 & 0.186 & 0.068 & 0.114 & 0.026 & 0.087 & 0.109 & 0     & 0.128 \\
5Y9J & 0     & 0.384 & 0     & 0.423 & 0.382 & 0     & 0.202 & 0.361 \\
6A0Z & 0.182 & 0.204 & 0.207 & 0.221 & 0.062 & 0.214 & 0.218 & 0.201 \\
6A77 & 0.046 & 0.581 & 0.037 & 0.591 & 0.591 & 0.59  & 0.583 & 0.589 \\
6AL0 & 0     & 0.338 & 0     & 0     & 0.424 & 0.331 & 0.345 & 0.329 \\
6B0S & 0.053 & 0.053 & 0.081 & 0     & 0.081 & 0.09  & 0.087 & 0.087\\
Mean ± Std & \underline{0.069±0.079}  & 0.26±0.26      & \textbf{0.049±0.054} & 0.261±0.286  & 0.365±0.297    & 0.123±0.158 & 0.372±0.3  & 0.292±0.236
\\
\bottomrule
\end{tabular*}
\end{table*}

\begin{table*}[!h]
\centering
\footnotesize
\caption{ Per-target and overall hit rates on the DBM55-AF2 dataset. AFM-AF2Rank refers to AF-Multimer-based AF2Rank. In the 'Best' column represents each target's best-possible Top-10 result, which is an upper limit of the hit rates. The ‘a/b/c’ values for each target represent the number of decoys among the top-10 ranked decoys that are classified as Acceptable or higher-quality (a), Medium or higher-quality (b), and High-quality (c). The quality is calculated according to the CAPRI criteria metric. The last row is a summary of all targets. For example, ‘13/9/2’ means that among the 15 targets, the corresponding method can select Acceptable or higher quality decoys for 13 targets. For all targets in summary, the best performance of selecting Acceptable or higher-quality models, Medium or higher-quality models, and High-quality models are marked in bold, respectively.}
\label{tab:bm55_hit}  
\begin{tabular*}{1.2\textwidth}{@{\extracolsep{0pt}}llllllllll}
\toprule
Target & TopoQA & ComplexQA\cite{complexqa} & DProQA\cite{dproqa}&AFM-AF2Rank  & GNN-DOVE\cite{complexqa} & GOAP\cite{complexqa}    & ZRANK2\cite{complexqa}  & TRScore\cite{complexqa} & Best\cite{complexqa}  \\
\midrule
3SE8   & 9/9/0      & 5/3/0     & 8/8/0&4/3/0   & 3/3/0    & 8/8/0   & 2/2/0   & 7/6/0   & 10/10/0 \\
3U7Y   & 2/2/1      & 2/2/1     & 2/2/1&0/0/0   & 2/2/1    & 2/2/1   & 2/2/1   & 1/1/1   & 2/2/1   \\
3WD5   & 8/8/0      & 4/4/0     & 10/8/0&7/5/0  & 0/0/0    & 10/8/0  & 6/4/0   & 5/3/0   & 10/10/0 \\
4ETQ   & 1/1/0      & 0/0/0     & 1/1/0&0/0/0   & 0/0/0    & 1/1/0   & 1/1/0   & 0/0/0   & 1/1/0   \\
4M5Z   & 10/10/1    & 10/10/1   & 10/10/1&10/10/1 & 10/10/0  & 10/10/1 & 10/10/1 & 10/10/1 & 10/10/1 \\
5CBA   & 10/10/2    & 10/10/0   & 10/10/1&10/10/1 & 10/10/3  & 10/10/2 & 10/10/4 & 10/10/2 & 10/10/6 \\
5GRJ   & 10/10/0    & 3/3/0     & 10/10/0&4/3/0 & 2/2/0    & 10/9/0  & 5/4/0   & 6/5/0   & 10/10/0 \\
5HGG   & 10/0/0     & 10/0/0    & 8/0/0&10/0/0   & 8/0/0    & 10/0/0  & 10/0/0  & 8/0/0   & 10/0/0  \\
5KOV   & 1/0/0      & 1/0/0     & 0/0/0 &1/0/0  & 1/0/0    & 1/0/0   & 0/0/0   & 1/0/0   & 2/0/0   \\
5WK3   & 0/0/0      & 2/0/0     & 0/0/0&3/0/0   & 1/0/0    & 0/0/0   & 3/0/0   & 0/0/0   & 3/0/0   \\
5Y9J   & 2/0/0      & 2/0/0     & 4/0/0&1/0/0   & 0/0/0    & 5/0/0   & 5/0/0   & 0/0/0   & 8/0/0   \\
6A0Z   & 0/0/0      & 3/0/0     & 0/0/0 &2/0/0  & 2/0/0    & 0/0/0   & 0/0/0   & 0/0/0   & 3/0/0   \\
6A77   & 7/7/0      & 0/0/0     & 7/7/0&2/0/0   & 0/0/0    & 3/3/0   & 4/4/0   & 3/3/0   & 8/8/0   \\
6AL0   & 10/2/0     & 6/2/0     & 9/2/0&6/2/0   & 6/0/0    & 9/0/0   & 9/0/0   & 8/0/0   & 10/2/0  \\
6B0S   & 10/10/0    & 10/10/0   & 10/10/0&10/10/0 & 10/10/0  & 10/10/0 & 10/10/0 & 10/10/0 & 10/10/0 \\
Summary    & \textbf{13}/\textbf{10}/\textbf{3}    & \textbf{13}/8/2    & 12/\textbf{10}/\textbf{3} &\textbf{13}/8/2& 11/6/2   & \textbf{13}/9/\textbf{3}  & \textbf{13}/9/\textbf{3}  & 11/8/\textbf{3}  & 15/10/3
\\
\bottomrule
\end{tabular*}
\end{table*}

\begin{table}[h!]
\small
\caption{The DockQ score ranking loss results on HAF2 dataset.The final row reports the mean and standard deviation (Std) of the ranking loss of different methods. Results for the best-performing methods are shown in bold, and results for the second-best methods are underlined. -: GOAP failed on these targets due to the sequence length. AFM-AF2Rank refers to AF-Multimer-based AF2Rank.}
\label{tab:haf2_loss}  
\begin{tabular*}{1.2\textwidth}{@{\extracolsep{\fill}}lllllllll@{\extracolsep{\fill}}}
\toprule
Target  & TopoQA  & ComplexQA\cite{complexqa} & DProQA\cite{dproqa}&AFM-AF2Rank & GNN-DOVE\cite{complexqa} & GOAP\cite{complexqa}  & ZRANK2\cite{complexqa} & TRScore\cite{complexqa}  \\
\midrule
7ALA    & 0.229          & 0.221     & 0.232&0.19  & 0.226    & 0.234 & 0.229       & 0.208    \\
7AMV    & 0.004          & 0.025     & 0.01&0.021   & 0.342    & -     & 0.02        & 0.013    \\
7AOH    & 0.045          & 0.235     & 0.066&0.287  & 0.354    & -     & 0.066       & 0.031     \\
7D3Y    & 0.011          & 0.295     & 0.326&0.322  & 0.295    & 0.266 & 0.314       & 0.325     \\
7D7F    & 0.471          & 0.466     & 0.471&0  & 0.471    & 0.471 & 0.47        & 0.343     \\
7KBR    & 0.074          & 0.12      & 0.068&0.069  & 0.068    & 0.104 & 0.025       & 0.152     \\
7LXT    & 0.057          & 0.008     & 0.586&0.05  & 0.295    & 0.052 & 0.043       & 0.054     \\
7MRW    & 0.025          & 0.59      & 0.085&0.024  & 0.598    & -     & 0.075       & 0.555     \\
7NKZ    & 0.014          & 0.002     & 0.164&0.075  & 0.192    & 0.012 & 0.019       & 0.113     \\
7O27    & 0.067          & 0.13      & 0.03&0.045   & 0.334    & 0.033 & 0.057       & 0.053     \\
7O28    & 0.031          & 0.094     & 0.029&0.042  & 0.244    & 0.027 & 0.027       & 0.244     \\
7OEL    & 0.035          & 0.189     & 0.062&0.013  & 0.21     & 0.189 & 0.157       & 0.232     \\
7OZN    & 0.49           & 0.281     & 0.409&0.491  & 0.457    & 0.412 & 0.489       & 0.491     \\
Mean ± Std & \textbf{0.119±0.163} & 0.204±0.169&0.195±0.185     &\underline{0.125±0.146}   & 0.314±0.132    & 0.18±0.156  & 0.153±0.164 & 0.216±0.167     \\
\bottomrule
\end{tabular*}
\end{table}

\begin{table*}[!h]
\footnotesize
\centering
\caption{ Per-target and overall hit rates on the HAF2 dataset. In the 'Best' column represents each target's best-possible Top-10 result, which is an upper limit of the hit rates. The ‘a/b/c’ values for each target represent the number of decoys among the top-10 ranked decoys that are classified as Acceptable or higher-quality (a), Medium or higher-quality (b), and High-quality (c). The quality is calculated according to the CAPRI criteria metric. The last row is a summary of all targets. For example, ‘10/9/4’ means that among the 13 targets, the corresponding method can select Acceptable or higher-quality decoys for 10 targets. For all targets in summary, the best performance of selecting Acceptable or higher-quality models, Medium or higher-quality models, and High-quality models are marked in bold, respectively. -: GOAP failed on these targets due to the sequence length. AFM-AF2Rank refers to AF-Multimer-based AF2Rank.}
\label{tab:haf2_hit}  
\begin{tabular*}{1.2\textwidth}{@{\extracolsep{\fill}}llllllllll@{\extracolsep{\fill}}}
\toprule
Target & TopoQA   & ComplexQA\cite{complexqa} & DProQA\cite{dproqa}&AFM-AF2Rank  & GNN-DOVE\cite{complexqa} & GOAP\cite{complexqa}     & ZRANK2\cite{complexqa}   & TRScore\cite{complexqa}  & Best\cite{complexqa}     \\
\midrule
7ALA   & 0/0/0    & 0/0/0     & 0/0/0&0/0/0    & 0/0/0    & 0/0/0    & 0/0/0    & 0/0/0    & 1/0/0    \\
7AMV   & 10/10/10 & 10/10/10  & 10/10/10&10/10/10 & 9/9/0    & -        & 10/10/10 & 10/10/7  & 10/10/10 \\
7AOH   & 10/10/10 & 10/10/9   & 10/10/10&10/10/6 & 9/9/0    & -        & 10/10/10 & 10/10/6  & 10/10/10 \\
7D3Y   & 4/0/0    & 1/0/0     & 2/0/0&1/0/0    & 0/0/0    & 2/0/0    & 2/0/0    & 2/0/0    & 10/0/0   \\
7D7F   & 0/0/0    & 2/0/0     & 0/0/0&5/0/0    & 0/0/0    & 1/0/0    & 4/0/0    & 0/0/0    & 5/0/0    \\
7KBR   & 10/10/10 & 10/10/10  & 10/10/10&10/10/10 & 10/10/9  & 10/10/10 & 10/10/10 & 10/10/10 & 10/10/10 \\
7LXT   & 5/5/0    & 3/3/0     & 1/1/0&10/10/0    & 1/0/0    & 8/8/0    & 8/8/0    & 10/10/0  & 10/10/0  \\
7MRW   & 10/9/0   & 0/0/0     & 5/4/0&10/10/0    & 0/0/0    & -        & 10/10/0  & 1/0/0    & 10/10/0  \\
7NKZ   & 10/10/10 & 10/10/10  & 10/10/2&10/10/10  & 10/9/9   & 10/10/10 & 10/10/10 & 10/10/10 & 10/10/10 \\
7O27   & 10/10/0  & 10/10/0   & 10/10/0&10/10/0  & 10/4/0   & 10/10/0  & 10/10/0  & 10/10/0  & 10/10/0  \\
7O28   & 10/10/0  & 10/10/0   & 10/10/0&10/10/0  & 10/10/0  & 10/10/0  & 10/10/0  & 10/10/0  & 10/10/0  \\
7OEL   & 10/10/0  & 10/10/0   & 10/10/0&10/10/0  & 10/10/0  & 10/9/0   & 10/10/0  & 10/9/0   & 10/10/0  \\
7OZN   & 0/0/0    & 2/0/0     & 0/0/0&0/0/0    & 0/0/0    & 0/0/0    & 0/0/0    & 0/0/0    & 10/2/0   \\
Summary    & 10/\textbf{9}/\textbf{4}   & \textbf{11}/8/\textbf{4}    & 10/\textbf{9}/\textbf{4}&\textbf{11}/\textbf{9}/\textbf{4}   & 8/7/3    & 8/6/2    & \textbf{11}/\textbf{9}/\textbf{4}   & 10/8/\textbf{4}   & 13/10/4  \\
\bottomrule
\end{tabular*}
\end{table*}

\begin{table}[h]
\small
\caption{DBM55-AF2 dataset ranking loss of TopoQA and DProQA with five different random seeds. The results of DProQA are from \cite{dproqa}. The number proceeding \_ is a specific random seed. ‘Mean' row represents the mean and standard deviation of the ranking losses for all targets. 'Multi\_mean' represents the average of the mean ranking losses from multiple experiments conducted with the same model but different random seeds.}
\label{tab:bm55_loss}  
\begin{tabular*}{1.4\textwidth}{@{\extracolsep{\fill}}lllllllllll@{\extracolsep{\fill}}}
\toprule
Target                  & TopoQA\_111 & TopoQA\_222 & TopoQA\_520 & TopoQA\_888 & TopoQA\_999 & DProQA\_111 & DProQA\_222 & DProQA\_520 & DProQA\_888 & DProQA\_999 \\
\midrule
6AL0 & 0.343       & 0           & 0.338       & 0.343       & 0.343       & 0.156       & 0           & 0.156       & 0.156       & 0.156       \\
3SE8 & 0.103       & 0.102       & 0.079       & 0.079       & 0           & 0.041       & 0.079       & 0.041       & 0.041       & 0.068       \\
5GRJ & 0.124       & 0.024       & 0.033       & 0.124       & 0.124       & 0.095       & 0.024       & 0.012       & 0.012       & 0.124       \\
6A77 & 0.062       & 0.046       & 0.062       & 0.062       & 0.062       & 0.062       & 0.037       & 0           & 0.037       & 0.062       \\
4M5Z & 0.248       & 0.269       & 0.248       & 0.248       & 0.248       & 0.242       & 0.015       & 0.015       & 0.015       & 0.251       \\
4ETQ & 0           & 0           & 0           & 0           & 0           & 0           & 0           & 0.75        & 0.748       & 0.75        \\
5CBA & 0           & 0.008       & 0.008       & 0.008       & 0.008       & 0.038       & 0.052       & 0.054       & 0.052       & 0.071       \\
5WK3 & 0.112       & 0.186       & 0.112       & 0.186       & 0.112       & 0.123       & 0.114       & 0.186       & 0.114       & 0.186       \\
5Y9J & 0           & 0           & 0           & 0           & 0           & 0           & 0           & 0           & 0           & 0           \\
6B0S & 0.053       & 0.053       & 0.053       & 0.053       & 0.053       & 0           & 0.081       & 0           & 0           & 0.081       \\
5HGG & 0.051       & 0.079       & 0.051       & 0.056       & 0.056       & 0.007       & 0.051       & 0.051       & 0.051       & 0.051       \\
6A0Z & 0.22        & 0.182       & 0.22        & 0.22        & 0.22        & 0.207       & 0.207       & 0.214       & 0.207       & 0.207       \\
3U7Y & 0           & 0.021       & 0           & 0           & 0           & 0           & 0           & 0           & 0           & 0.021       \\
3WD5 & 0.011       & 0.011       & 0           & 0.011       & 0           & 0.109       & 0.011       & 0           & 0.192       & 0.109       \\
5KOV & 0.087       & 0.06        & 0.07        & 0.06        & 0.083       & 0.087       & 0.065       & 0.087       & 0.087       & 0.09        \\
Mean & 0.094±0.100 & 0.069±0.079 & 0.085±0.100 & 0.097±0.102 & 0.087±0.103 & 0.078±0.076 & 0.049±0.054 & 0.104±0.186 & 0.114±0.182 & 0.148±0.174 \\
\cline{2-6}\cline{7-11}
Multi\_mean & \multicolumn{5}{c}{0.087±0.010}                                     & \multicolumn{5}{c}{0.099±0.033}                        
\\
\bottomrule
\end{tabular*}
\end{table}

\begin{table*}[!h]
\small
\caption{DBM55-AF2 dataset hit rate of TopoQA and DProQA with five different random seeds. The results of DProQA are from \cite{dproqa}. The number proceeding \_ is a specific random seed. The ‘Summary' row is a summary of all targets. For example, ‘13/9/2’ means that among the 15 targets, the corresponding method can select Acceptable or higher quality decoys for 13 targets. 'Multi\_mean' represents the average of the hit rate from multiple experiments conducted with the same model but different random seeds.}
\label{tab:bm55_loss}  
\begin{tabular*}{1.4\textwidth}{@{\extracolsep{\fill}}lllllllllll@{\extracolsep{\fill}}}

\toprule
Target                  & TopoQA\_111 & TopoQA\_222 & TopoQA\_520 & TopoQA\_888 & TopoQA\_999 & DPROQ\_111 & DPROQ\_222 & DPROQ\_520 & DPROQ\_888 & DPROQ\_999 \\
\midrule
6AL0    & 10/2/0  & 10/2/0  & 9/0/0   & 10/2/0  & 10/2/0  & 10/2/0  & 9/2/0   & 10/2/0  & 10/2/0  & 10/1/0  \\
3SE8    & 9/9/0   & 9/9/0   & 9/9/0   & 10/10/0 & 10/10/0 & 10/10/0 & 8/8/0   & 8/8/0   & 10/10/0 & 8/8/0   \\
5GRJ    & 10/10/0 & 10/10/0 & 10/10/0 & 10/10/0 & 10/10/0 & 10/10/0 & 10/10/0 & 9/9/0   & 9/9/0   & 9/9/0   \\
6A77    & 7/7/0   & 7/7/0   & 7/7/0   & 7/7/0   & 7/7/0   & 7/7/0   & 7/7/0   & 8/8/0   & 7/7/0   & 8/8/0   \\
4M5Z    & 10/10/0 & 10/10/1 & 10/10/1 & 10/10/0 & 10/10/1 & 10/10/1 & 10/10/1 & 10/10/0 & 10/10/1 & 10/10/0 \\
4ETQ    & 1/1/0   & 1/1/0   & 1/1/0   & 1/1/0   & 1/1/0   & 1/1/0   & 1/1/0   & 1/1/0   & 1/1/0   & 1/1/0   \\
5CBA    & 10/10/2 & 10/10/2 & 10/10/4 & 10/10/2 & 10/10/2 & 10/10/0 & 10/10/1 & 10/10/1 & 10/10/1 & 10/10/1 \\
5WK3    & 2/0/0   & 0/0/0   & 0/0/0   & 0/0/0   & 2/0/0   & 0/0/0   & 0/0/0   & 0/0/0   & 0/0/0   & 0/0/0   \\
5Y9J    & 8/0/0   & 2/0/0   & 6/0/0   & 2/0/0   & 7/0/0   & 6/0/0   & 4/0/0   & 8/0/0   & 5/0/0   & 4/0/0   \\
6B0S    & 10/10/0 & 10/10/0 & 10/10/0 & 10/10/0 & 10/10/0 & 10/10/0 & 10/10/0 & 10/10/0 & 10/10/0 & 10/10/0 \\
5HGG    & 10/0/0  & 10/0/0  & 10/0/0  & 10/0/0  & 10/0/0  & 8/0/0   & 8/0/0   & 8/0/0   & 8/0/0   & 8/0/0   \\
6A0Z    & 1/0/0   & 0/0/0   & 1/0/0   & 1/0/0   & 1/0/0   & 0/0/0   & 0/0/0   & 0/0/0   & 0/0/0   & 0/0/0   \\
3U7Y    & 2/2/1   & 2/2/1   & 2/2/1   & 2/2/1   & 2/2/1   & 2/2/1   & 2/2/1   & 2/2/1   & 2/2/1   & 2/2/1   \\
3WD5    & 9/8/0   & 8/8/0   & 9/9/0   & 9/8/0   & 9/8/0   & 9/8/0   & 10/8/0  & 8/8/0   & 9/8/0   & 9/8/0   \\
5KOV    & 1/0/0   & 1/0/0   & 1/0/0   & 1/0/0   & 1/0/0   & 0/0/0   & 0/0/0   & 0/0/0   & 0/0/0   & 0/0/0   \\
Summary & 15/10/2 & 13/10/3 & 14/9/3  & 14/10/2 & 15/10/3 & 12/10/2 & 12/10/3 & 12/10/2 & 12/10/3 & 12/10/2\\
\cline{2-6}\cline{7-11}
Multi\_mean & \multicolumn{5}{c}{14.2/9.8/2.6}                                      & \multicolumn{5}{c}{12/10/2.4}      
\\
\bottomrule
\end{tabular*}
\end{table*}

\begin{table*}[!th]
\small
\caption{HAF2 dataset ranking loss of TopoQA and DProQA with five different random seeds. The results of DProQA are from \cite{dproqa}. The number proceeding \_ is a specific random seed. ‘Mean' row represents the mean and standard deviation of the ranking losses for all targets. 'Multi\_mean' represents the average of the mean ranking losses from multiple experiments conducted with the same model but different random seeds.}
\label{tab:haf2_loss}  
\begin{tabular*}{1.4\textwidth}{@{\extracolsep{\fill}}lllllllllll@{\extracolsep{\fill}}}
\toprule
Target                  & TopoQA\_111 & TopoQA\_222 & TopoQA\_520 & TopoQA\_888 & TopoQA\_999 & DPROQ\_111  & DPROQ\_222  & DPROQ\_520  & DPROQ\_888  & DPROQ\_999  \\
\midrule
7AOH & 0.005       & 0.045       & 0.061       & 0.045       & 0.046       & 0.05        & 0.066       & 0.026       & 0.026       & 0.066       \\
7D7F & 0.471       & 0.471       & 0.471       & 0.471       & 0.471       & 0.471       & 0.471       & 0.47        & 0.47        & 0.471       \\
7AMV & 0.007       & 0.004       & 0.007       & 0.065       & 0.007       & 0.019       & 0.01        & 0.019       & 0.019       & 0.017       \\
7OEL & 0.035       & 0.035       & 0.035       & 0.035       & 0.035       & 0.135       & 0.062       & 0.063       & 0.135       & 0.063       \\
7O28 & 0.031       & 0.031       & 0.057       & 0.031       & 0.031       & 0.021       & 0.029       & 0.151       & 0.021       & 0.02        \\
7ALA & 0.233       & 0.229       & 0.229       & 0.229       & 0.233       & 0.226       & 0.232       & 0.234       & 0.227       & 0.234       \\
7MRW & 0.024       & 0.025       & 0.014       & 0.117       & 0.014       & 0.555       & 0.085       & 0.599       & 0.555       & 0.555       \\
7OZN & 0.49        & 0.49        & 0.49        & 0.493       & 0.412       & 0.412       & 0.409       & 0.493       & 0.412       & 0.409       \\
7D3Y & 0.023       & 0.011       & 0.03        & 0.011       & 0.011       & 0.326       & 0.326       & 0.326       & 0.326       & 0.326       \\
7NKZ & 0.008       & 0.014       & 0.008       & 0.008       & 0           & 0.164       & 0.164       & 0.175       & 0.164       & 0.164       \\
7LXT & 0.057       & 0.057       & 0.054       & 0.057       & 0.057       & 0.586       & 0.586       & 0.586       & 0.586       & 0.586       \\
7KBR & 0.064       & 0.074       & 0           & 0.12        & 0.12        & 0.193       & 0.068       & 0.095       & 0.026       & 0.152       \\
7O27 & 0.019       & 0.067       & 0.067       & 0.061       & 0.077       & 0.03        & 0.03        & 0.079       & 0.079       & 0.03        \\
Mean & 0.113±0.167 & 0.112±0.163 & 0.117±0.165 & 0.134±0.159 & 0.116±0.151 & 0.245±0.197 & 0.195±0.185 & 0.255±0.207 & 0.234±0.204 & 0.238±0.201 \\
\cline{2-6}\cline{7-11}
Multi\_mean & \multicolumn{5}{c}{0.120±0.006}                                     & \multicolumn{5}{c}{0.233±0.020}  
\\
\bottomrule
\end{tabular*}
\end{table*}

\begin{table}[!h]
\small
\caption{HAF2 dataset hit rate of TopoQA and DProQA with five different random seeds. The results of DProQA are from \cite{dproqa}. The number proceeding \_ is a specific random seed. The ‘Summary' row is a summary of all targets. For example, ‘10/9/4’ means that among the 13 targets, the corresponding method can select Acceptable or higher quality decoys for 10 targets. 'Multi\_mean' represents the average of the hit rate from multiple experiments conducted with the same model but different random seeds.}
\label{tab:haf2_hit}  
\begin{tabular*}{1.4\textwidth}{@{\extracolsep{\fill}}lllllllllll@{\extracolsep{\fill}}}

\toprule
Target                  & TopoQA\_111 & TopoQA\_222 & TopoQA\_520 & TopoQA\_888 & TopoQA\_999 & DPROQ\_111 & DPROQ\_222 & DPROQ\_520 & DPROQ\_888 & DPROQ\_999 \\
\midrule
7AOH & 10/10/10 & 10/10/10 & 10/10/10 & 10/10/10 & 10/10/10 & 10/10/10 & 10/10/10 & 10/10/10 & 10/10/1 & 10/10/1 \\
7D7F & 0/0/0    & 0/0/0    & 0/0/0    & 0/0/0    & 0/0/0    & 0/0/0    & 0/0/0    & 0/0/0    & 0/0/0   & 0/0/0   \\
7AMV & 10/10/10 & 10/10/10 & 10/10/10 & 10/10/10 & 10/10/10 & 10/10/10 & 10/10/10 & 10/10/10 & 10/10/1 & 10/10/1 \\
7OEL & 10/10/0  & 10/10/0  & 10/10/0  & 10/10/0  & 10/9/0   & 10/10/0  & 10/10/0  & 10/10/0  & 10/10/0 & 10/10/0 \\
7O28 & 10/10/0  & 10/10/0  & 10/10/0  & 10/10/0  & 10/10/0  & 10/10/0  & 10/10/0  & 10/10/0  & 10/10/0 & 10/10/0 \\
7ALA & 0/0/0    & 0/0/0    & 0/0/0    & 0/0/0    & 0/0/0    & 0/0/0    & 0/0/0    & 0/0/0    & 0/0/0   & 0/0/0   \\
7MRW & 10/10/0  & 10/9/0   & 10/10/0  & 10/10/0  & 10/10/0  & 0/0/0    & 5/4/0    & 0/0/0    & 0/0/0   & 2/2/0   \\
7OZN & 0/0/0    & 0/0/0    & 0/0/0    & 0/0/0    & 0/0/0    & 0/0/0    & 0/0/0    & 0/0/0    & 0/0/0   & 0/0/0   \\
7D3Y & 6/0/0    & 4/0/0    & 9/0/0    & 5/0/0    & 5/0/0    & 6/0/0    & 2/0/0    & 7/0/0    & 4/0/0   & 8/0/0   \\
7NKZ & 10/10/9  & 10/10/10 & 10/10/9  & 10/10/10 & 10/10/10 & 10/10/2  & 10/10/2  & 10/10/1  & 10/10/1 & 10/10/2 \\
7LXT & 4/4/0    & 5/5/0    & 4/4/0    & 4/4/0    & 7/7/0    & 0/0/0    & 1/1/0    & 0/0/0    & 0/0/0   & 0/0/0   \\
7KBR & 10/10/10 & 10/10/10 & 10/10/10 & 10/10/10 & 10/10/10 & 10/10/-  & 10/10/10 & 10/10/10 & 10/10/9 & 10/10/9 \\
7O27 & 10/10/0  & 10/10/0  & 10/10/0  & 10/9/0   & 10/10/0  & 10/10/0  & 10/10/0  & 10/10/0  & 10/10/0 & 10/10/0 \\
Summary                 & 10/9/4      & 10/9/4      & 10/9/4      & 10/9/4      & 10/9/4      & 8/7/4      & 10/9/4     & 8/7/4      & 9/8/4      & 8/7/3      \\
\cline{2-6}\cline{7-11}
Multi\_mean & \multicolumn{5}{c}{10/9/4}                                          & \multicolumn{5}{c}{8.6/7.6/4} 
\\
\bottomrule
\end{tabular*}
\end{table}

\end{landscape}
\renewcommand{\thetable}{S\arabic{table}}
\begin{table}[!t]
\centering
\caption{The DockQ-wave score ranking loss results on ABAG-AF3 dataset. ‘Mean ± Std' row represents the mean and standard deviation of the ranking losses for all targets. -: AF-Multimer failed on these targets due to out of memory. We ran the trained models of ComplexQA and DProQA from GitHub to obtain results. The AF3 results come from the output self-assessment scores. AFM-AF2Rank refers to AF-Multimer-based AF2Rank, and its results were obtained by running AF2Rank, which is accessible at: https://colab.research.google.com/github/sokrypton/ColabDesign/blob/main/af/examples/AF2Rank.ipynb.}
\label{tab:haf2_hit}  
\begin{tabular*}{\textwidth}{@{\extracolsep{\fill}}llllll@{\extracolsep{\fill}}}
\toprule
Target     & TopoQA      & AF3         & DProQA      & ComplexQA   & AFM-AF2Rank           \\
\midrule
7Z4T       & 0.011       & 0.006       & 0.307       & 0.008       & 0.002                 \\
7OM4       & 0.203       & 0.087       & 0.228       & 0.203       & 0.237                 \\
7R40       & 0.000       & 0.066       & 0.014       & 0.002       & 0.006                 \\
7SBG       & 0.028       & 0.029       & 0.256       & 0.205       & 0.043                 \\
7O9W       & 0.048       & 0.051       & 0.039       & 0.126       & 0.039                 \\
7YQZ       & 0.019       & 0.029       & 0.059       & 0.000       & - \\
7T77       & 0.018       & 0.006       & 0.016       & 0.018       & 0.002                 \\
7S0E       & 0.005       & 0.009       & 0.006       & 0.018       & 0.014                 \\
7YQX       & 0.013       & 0.021       & 0.017       & 0.021       & - \\
8GV6       & 0.176       & 0.179       & 0.131       & 0.247       & 0.235                 \\
7VYR       & 0.049       & 0.034       & 0.046       & 0.046       & 0.026                 \\
7SU0       & 0.275       & 0.010       & 0.281       & 0.288       & 0.000                 \\
7VNG       & 0.020       & 0.100       & 0.332       & 0.194       & 0.316                 \\
7ZR7       & 0.332       & 0.305       & 0.337       & 0.406       & 0.013                 \\
8GZ5       & 0.380       & 0.018       & 0.823       & 0.000       & 0.824                 \\
7TFO       & 0.032       & 0.022       & 0.034       & 0.022       & 0.022                 \\
7SD3       & 0.003       & 0.016       & 0.051       & 0.061       & 0.038                 \\
7SGM       & 0.275       & 0.006       & 0.068       & 0.278       & 0.068                 \\
7X7O       & 0.001       & 0.007       & 0.001       & 0.015       & 0.009                 \\
7SJO       & 0.050       & 0.050       & 0.059       & 0.061       & 0.050                 \\
7TEE       & 0.089       & 0.162       & 0.162       & 0.163       & 0.166                 \\
7UED       & 0.017       & 0.020       & 0.036       & 0.045       & 0.029                 \\
8B7W       & 0.000       & 0.187       & 0.158       & 0.158       & 0.187                 \\
7SU1       & 0.019       & 0.013       & 0.010       & 0.002       & 0.014                 \\
7T25       & 0.005       & 0.006       & 0.001       & 0.209       & 0.001                 \\
8F8X       & 0.094       & 0.091       & 0.085       & 0.093       & 0.091                 \\
8GV7       & 0.000       & 0.018       & 0.051       & 0.054       & 0.037                 \\
7Z2M       & 0.132       & 0.140       & 0.119       & 0.125       & 0.124                 \\
7TYV       & 0.031       & 0.021       & 0.129       & 0.136       & 0.037                 \\
7T73       & 0.071       & 0.070       & 0.055       & 0.081       & 0.072                 \\
7WO5       & 0.238       & 0.000       & 0.230       & 0.094       & 0.151                 \\
7ZF9       & 0.031       & 0.030       & 0.067       & 0.000       & 0.023                 \\
7WRV       & 0.200       & 0.073       & 0.130       & 0.189       & 0.203                 \\
7SJN       & 0.338       & 0.010       & 0.007       & 0.007       & 0.012                 \\
7SWN       & 0.009       & 0.010       & 0.000       & 0.116       & 0.013                 \\
Mean ± Std & \underline{0.092±0.113} & \textbf{0.054±0.067} & 0.124±0.157 & 0.106±0.100 & 0.094±0.153\\    
\bottomrule
\end{tabular*}
\end{table}

\begin{table}[!t]
\centering
\caption{The Top-10 mean DockQ-wave results on ABAG-AF3 dataset. The 'Mean' row indicates the average of the mean DockQ-wave values calculated from the top 10 models of each target, and then averaged over all targets. -: AF-Multimer failed on these targets due to out of memory. We ran the trained models of ComplexQA and DProQA from GitHub to obtain results. The AF3 results come from the output self-assessment scores. AFM-AF2Rank refers to AF-Multimer-based AF2Rank, and its results were obtained by running AF2Rank, which is accessible at: https://colab.research.google.com/github/sokrypton/ColabDesign/blob/main/af/examples/AF2Rank.ipynb.}
\label{tab:haf2_hit}  
\begin{tabular*}{\textwidth}{@{\extracolsep{\fill}}llllll@{\extracolsep{\fill}}}
\toprule
Target & TopoQA & AF3    & DProQA & ComplexQA & AFM-AF2Rank \\
\midrule
7Z4T   & 0.7929 & 0.8530 & 0.7357 & 0.8220    & 0.7927      \\
7OM4   & 0.5156 & 0.5427 & 0.5170 & 0.4665    & 0.5363      \\
7R40   & 0.6000 & 0.5969 & 0.6007 & 0.6242    & 0.5762      \\
7SBG   & 0.6325 & 0.6553 & 0.4793 & 0.5759    & 0.6608      \\
7O9W   & 0.7053 & 0.6911 & 0.7078 & 0.6849    & 0.6980      \\
7YQZ   & 0.4442 & 0.4399 & 0.4253 & 0.4360    & -           \\
7T77   & 0.5974 & 0.6006 & 0.5992 & 0.6007    & 0.5998      \\
7S0E   & 0.5041 & 0.5037 & 0.5033 & 0.5047    & 0.5056      \\
7YQX   & 0.4362 & 0.4466 & 0.4292 & 0.4426    & -           \\
8GV6   & 0.5030 & 0.5822 & 0.5353 & 0.5101    & 0.4724      \\
7VYR   & 0.8460 & 0.8305 & 0.8190 & 0.8418    & 0.8358      \\
7SU0   & 0.7341 & 0.7884 & 0.6588 & 0.7750    & 0.7925      \\
7VNG   & 0.6836 & 0.7103 & 0.5708 & 0.6316    & 0.6106      \\
7ZR7   & 0.5538 & 0.5457 & 0.5506 & 0.6335    & 0.5889      \\
8GZ5   & 0.7548 & 0.8669 & 0.6986 & 0.7887    & 0.1705      \\
7TFO   & 0.4550 & 0.4574 & 0.4541 & 0.4542    & 0.4551      \\
7SD3   & 0.5753 & 0.5875 & 0.5824 & 0.5748    & 0.5720      \\
7SGM   & 0.6637 & 0.6853 & 0.7220 & 0.5859    & 0.7123      \\
7X7O   & 0.7914 & 0.7859 & 0.7885 & 0.7902    & 0.7869      \\
7SJO   & 0.5812 & 0.5790 & 0.5806 & 0.5884    & 0.5792      \\
7TEE   & 0.4842 & 0.4655 & 0.4847 & 0.4151    & 0.4132      \\
7UED   & 0.7653 & 0.7648 & 0.7614 & 0.7288    & 0.7619      \\
8B7W   & 0.0896 & 0.0744 & 0.0697 & 0.0681    & 0.0433      \\
7SU1   & 0.8481 & 0.8479 & 0.8477 & 0.8480    & 0.8494      \\
7T25   & 0.7836 & 0.8211 & 0.8214 & 0.7463    & 0.8025      \\
8F8X   & 0.0281 & 0.0110 & 0.0108 & 0.0120    & 0.0225      \\
8GV7   & 0.5912 & 0.5941 & 0.5826 & 0.5719    & 0.5871      \\
7Z2M   & 0.4364 & 0.4192 & 0.4257 & 0.4501    & 0.4484      \\
7TYV   & 0.7460 & 0.7212 & 0.7091 & 0.6786    & 0.7438      \\
7T73   & 0.6083 & 0.6083 & 0.6126 & 0.6109    & 0.6067      \\
7WO5   & 0.4350 & 0.4984 & 0.4543 & 0.4604    & 0.4910      \\
7ZF9   & 0.8891 & 0.8807 & 0.8735 & 0.8858    & 0.8689      \\
7WRV   & 0.3538 & 0.4247 & 0.3517 & 0.3520    & 0.3286      \\
7SJN   & 0.5956 & 0.8907 & 0.8907 & 0.8611    & 0.8574      \\
7SWN   & 0.7020 & 0.7183 & 0.6340 & 0.6328    & 0.6732      \\
Mean   & \underline{0.592}  & \textbf{0.614}  & 0.585  & 0.590     & 0.589 \\ 
\bottomrule
\end{tabular*}
\end{table}

\renewcommand{\thefigure}{S\arabic{figure}}
\setcounter{figure}{0}  

\begin{table*}[t]
\centering
\small
\caption{Ablation study results on DBM55-AF2 dataset and HAF2 dataset.In the ablation studies, we used DockQ as the reference metric. For each metric in each dataset, we bolded the best results.}
\label{tab:ablation}  
\begin{tabular*}{\textwidth}{@{\extracolsep{\fill}}lllllll@{\extracolsep{\fill}}}
\toprule
Method                 & \multicolumn{3}{l}{DBM55-AF2}             & \multicolumn{3}{l}{HAF2}       \\
\cline{2-4}\cline{5-7}
& Ranking loss & PearsonCor & SpearCor & Ranking loss & PearsonCor & SpearCor  \\
\midrule
w/o node topological features & 0.129        & 0.317      & 0.38     & 0.157        & 0.167      & 0.266     \\
w/o edge features related to atomic distances & 0.103        & \textbf{0.525}      & 0.491    & 0.165       & 0.629     & 0.634    \\
TopoQA                                   & \textbf{0.069}        & 0.515      & \textbf{0.502}    & \textbf{0.119}        & \textbf{0.649}      & \textbf{0.683}     \\
\bottomrule
\end{tabular*}
\end{table*}
\clearpage





\begin{figure}[!h]%
\centering
\includegraphics[width=\textwidth]{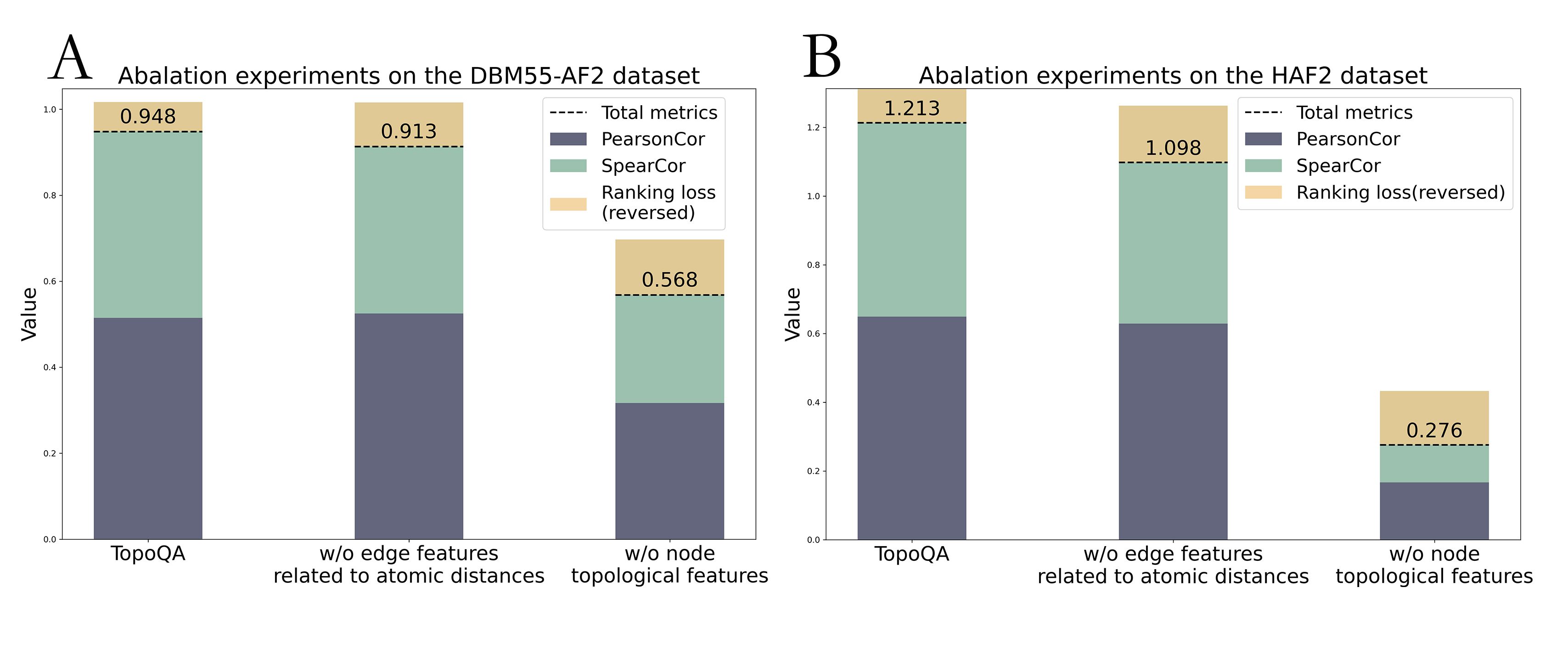}
\caption{Results of ablation study. (A) Abalation experiments on the DBM55-AF2 dataset. (B) Abalation experiments on the HAF2 dataset. For Pearson and Spearman correlation coefficients, higher values are better, while for ranking loss, lower values are better. We use the negative values of ranking loss for plotting to ensure consistency across all metrics. The length of each small square is proportional to the metric's value: positive values are plotted upwards, while negative values are plotted downwards. The numbers displayed above the bars for each method represent the sum of all metrics.}\label{abalation}
\end{figure}
\newpage

\begin{figure}[!t]%
\centering
\includegraphics[width=0.8\textwidth]{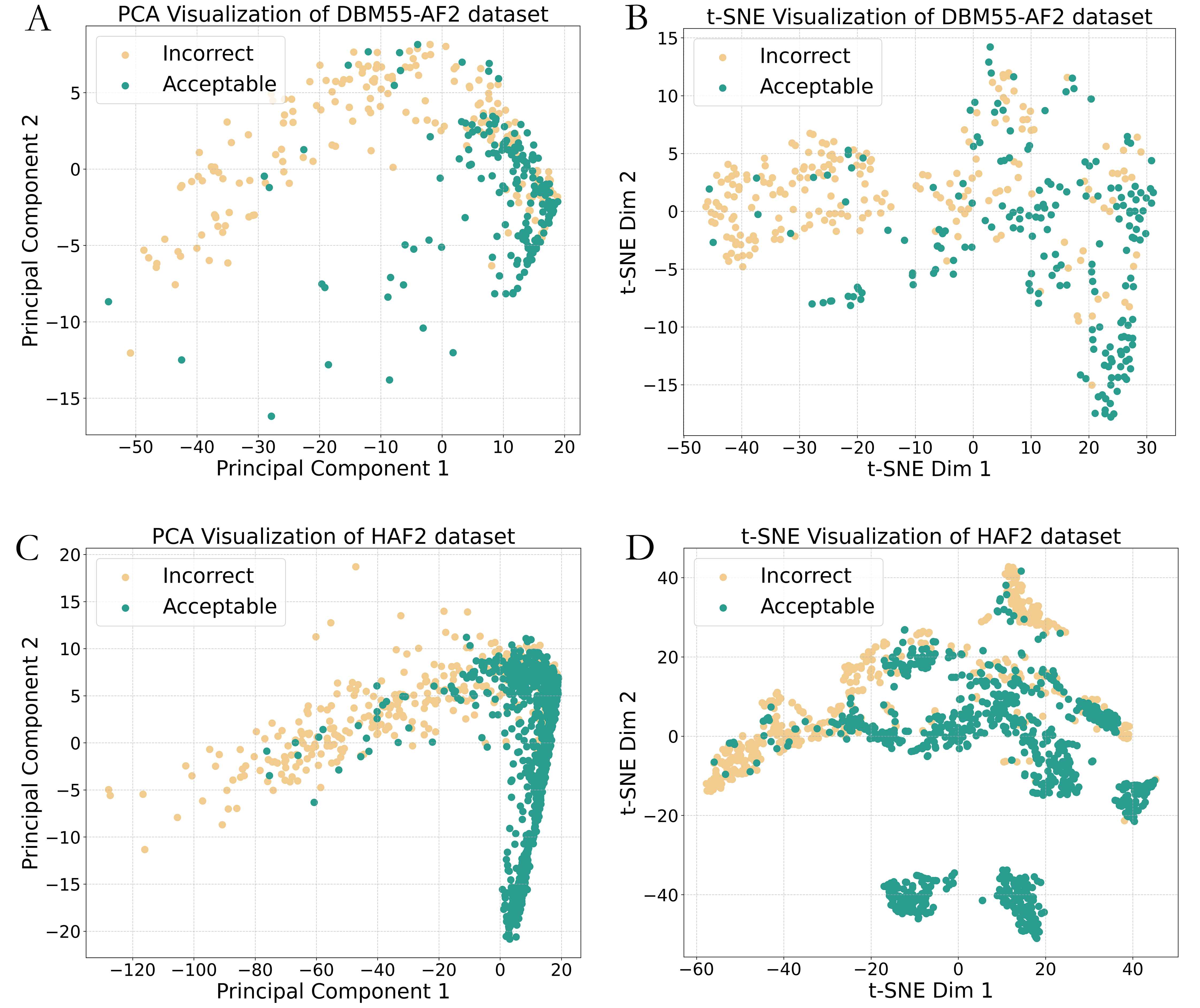}
\caption{Feature visualization by PCA and t-SNE. (A) PCA visualization of DBM55-AF2 dataset. (B) t-SNE visualization of DBM55-AF2 dataset.(C) PCA visualization of HAF2 dataset. (D) t-SNE visualization of HAF2 dataset.}\label{keshihua}
\end{figure}

\end{document}